\documentclass[%
 aip,
 cha,%
 amsmath,amssymb,
 reprint,%
]{revtex4-1}

\usepackage{graphicx}
\usepackage{dcolumn}
\usepackage{bm}

\usepackage{algorithm}
\usepackage{algpseudocode}
\usepackage{caption}
\usepackage{subcaption}
\graphicspath{{Figures/}}

\usepackage{array}
\newcolumntype{x}[1]{%
>{\centering\hspace{0pt}}p{#1}}%

\usepackage{color}

\usepackage{pifont}
\newcommand{\vmark}{\ding{52}}%
\newcommand{\xmark}{\ding{54}}%

\begin{document}

\preprint{AIP/123-QED}

\title[Complex contagions with timers]{Complex contagions with timers
}

\author{Se-Wook Oh}
\affiliation{Oxford Centre for Industrial and Applied Mathematics, Mathematical Institute, University of Oxford, Oxford OX2 6GG, UK}

\author{Mason A. Porter}
\affiliation{Department of Mathematics, University of Los Angeles, Los Angeles, USA}
\affiliation{Oxford Centre for Industrial and Applied Mathematics, Mathematical Institute, University of Oxford, Oxford OX2 6GG, UK}
\affiliation{CABDyN Complexity Centre, University of Oxford, Oxford OX1 1HP, UK}

\date{\today}

\begin{abstract}
A great deal of effort has gone into trying to model social influence --- including the spread of behavior, norms, and ideas --- on networks. Most models of social influence tend to assume that individuals react to changes in the states of their neighbors without any time delay, but this is often not true in social contexts, where (for various reasons) different agents can have different response times. To examine such situations, we introduce the idea of a timer into threshold models of social influence. The presence of timers on nodes delays the adoption --- i.e., change of state --- of each agent, which in turn delays the adoptions of its neighbors. With a homogeneous-distributed timer, in which all nodes exhibit the same amount of delay, adoption delays are also homogeneous, so the adoption order of nodes remains the same. However, heterogeneously-distributed timers can change the adoption order of nodes and hence the ``adoption paths'' through which state changes spread in a network. Using a threshold model of social contagions, we illustrate that heterogeneous timers can either accelerate or decelerate the spread of adoptions compared to an analogous situation with homogeneous timers, and we investigate the relationship of such acceleration or deceleration with respect to timer distribution and network structure. We derive an analytical approximation for the temporal evolution of the fraction of adopters by modifying a pair approximation of the Watts threshold model, and we find good agreement with numerical computations. We also examine our new timer model on networks constructed from empirical data.
\end{abstract}

\maketitle

\begin{quotation}

Mathematical modeling of social contagions is a useful framework for studying the spread of phenomena such as ideas, memes, misinformation, and ``alternative facts'' on networks \cite{rogers2010diffusion,yamir-jcn2013,porter2016dynamical}. In most models, including classical threshold models \cite{valente1996social,watts2002simple,kempe2003maximizing} and their generalizations, the rules for updating node states depend only on the states of the nodes' nearest neighbors. We introduce a temporal element into such update rules by incorporating a \emph{timer} into the adoption condition to model the tendency of individuals to wait some amount of time before they adopt behavior from their neighbors. This idea is relevant for numerous models for social contagions (and other spreading processes), but for concreteness we incorporate timers into the popular Watts threshold model (WTM) \cite{valente1996social,watts2002simple} of social influence. In our model, each node has both a threshold and a timer; once its threshold is matched or surpassed (by the fraction of adopted nodes in its neighborhood being at least as large as this threshold), a countdown begins, and the node changes its state to adopt this behavior once the timer reaches $0$. We investigate the dynamics of the WTM with a timer for both homogeneously-distributed and heterogeneously-distributed timers, and we derive an analytical approximation using a pair approximation that gives good agreement with numerical computations for the temporal evolution of adoptions.
 
\end{quotation}



\section{Introduction} \label{sec:level1}

Over the decades, and especially recently amidst the surge in data availability and richness, it has become increasingly popular to take quantitative approaches to the study of sociological questions \cite{science2009,borgatti2009network,margetts2015political,watts2017,hofman2017}. Modeling efforts have drawn from mathematical, statistical, and computational approaches \cite{porter2017}; and the study of mechanistic models that incorporate data in a meaningful way (see [\onlinecite{gleeson2014simple}] for an example of data-driven mechanistic modeling) can give insight into both existing observations and forecasting of future dynamics. For example, there have been numerous studies of the spread of opinions, actions, memes, information, misinformation, ``alternative facts'', and other phenomena in populations in disciplines such as sociology, economics, computer science, physics, and many others\cite{porter2016dynamical,fowlerreview,yamir-jcn2013,jackson2014,young2009}. By analogy with the spread of infectious diseases in a population, spreading phenomena --- including the spread of defaults of banks, norms in populations, and products or new practices in populations --- are often modeled as contagion processes on a network. To distinguish between different mechanisms in social and biological contagions, the former are often construed as examples of ``complex contagions'', and the latter are often construed as examples of ``simple contagions''\cite{weng2013virality,lerman2016information,monsted2017}.

The quantitative study of the spread of behavior in social networks has a long history that dates back (at least) several decades \cite{Bass69,schelling1971dynamic,schelling1973hockey,friedkinbook,granovetter1978threshold,valente1996social}. In recent years, studies of social influence have tended to focus on large social and/or communication networks, taking advantage of the increased availability of microblogging data sets (e.g., using data from Twitter) with relational information that enables one to incorporate network effects into models\cite{gonzalez2011dynamics, piedrahita2013modeling}. In studying social influence, one often explores what conditions yield \emph{cascades}\cite{motter2017}, in which a small seed of activity leads to a large change in a network. In the study of spreading models, a common way to quantify cascades is to examine when an infinitesimally small seed fraction of adopted nodes generates a nonvanishing mean cascade size as the total number $N$ of nodes in a network becomes infinite ($N\rightarrow\infty$)\cite{gleeson2008cascades,porter2016dynamical}. In practical applications with empirical data, one often measures cascade sizes in other ways (such as by calculating how long it takes for a given fraction of the nodes in a network to adopt). See the discussion of cascade conditions in [\onlinecite{porter2016dynamical}].

A particularly popular framework for studying the spread of behavior in social networks are \emph{threshold models}, in which nodes update their states if the amount of peer pressure from their neighboring nodes (usually just nearest neighbors) exceeds some personal threshold. The simplest such model is the Watts threshold model (WTM)\cite{watts2002simple}, which uses a linear updating rule that is very similar
\footnote{In [\protect\onlinecite{granovetter1978threshold}], which does not incorporate network structure (it assumes that each node is adjacent to all other nodes), Granovetter did not define the adoption threshold of a node in terms of either the fraction or the total number of its adopted neighbors. Because of the all-to-all connectivity in [\protect\onlinecite{granovetter1978threshold}], Granovetter's linear-threshold update rule does not distinguish between the fraction (as in [\protect\onlinecite{watts2002simple}]) and number (as in [\protect\onlinecite{centola2007complex}]) of adopted neighbors. There is also a slight difference between Granovetter's update rule and the one in the WTM: in the former, a node will not adopt under any circumstance if its threshold is $1$; in the latter, a node with threshold $1$ adopts if all of its neighbors have adopted. Finally, although threshold models are used to model social contagions (and for so-called ``complex contagions''), Granovetter wrote in [\protect\onlinecite{granovetter1978threshold}] that ``In the broader context of threshold models, the idea of `contagion' seems inappropriate, since much more is involved than mere imitation of the last person observed.''} to the one introduced by Granovetter \cite{granovetter1978threshold} and which was previously examined on networks by Valente \cite{valente95,valente1996social}. (It is also related to bootstrap percolation \cite{chalupa1979}.) The WTM uses a ``threshold'' to represent the latent tendency of an individual to adopt an innovation (or become infected, if one wants to use more biological terminology) when at least some fraction of its neighbors has adopted the innovation\cite{kempe2003maximizing}. 
Threshold models of adoption were first studied by sociologists, and the idea of a threshold comes from a sociological theory\cite{granovetter1978threshold} that articulates that a person exhibits inertia in adopting an innovation as a way to reduce cost in making decisions. The WTM is appealing to study in part because of its mathematical tractability \cite{gleeson2007seed, gleeson2013binary,porter2016dynamical} and in part because it incorporates simple notions of peer pressure, social reinforcement (because multiple neighbors who have adopted an idea increases the peer pressure for a node to adopt), and personal resistance to peer pressure.\footnote{In some applications, one sometimes uses terms like ``infected'' or ``active'' rather than ``adopted''.} It thereby provides a simple model of social influence, which occurs when an individual is influenced by others and adopts their behavior\cite{robins2001network, leenders1997longitudinal}. 

The WTM and its generalizations have been studied from many perspectives. This includes a considerable amount of work --- both analytical and numerical --- on WTM dynamics on random networks with various characteristics, including local clustering \cite{hackett2011cascades, hackett2013cascades, mcsweeney2015single}, community structure \cite{gleeson2008cascades}, degree--degree correlations\cite{dodds2009analysis}, and communities with intercommunity correlations\cite{melnik2014dynamics}. The WTM has also been generalized to dynamics on temporal networks\cite{karimi2013threshold} and multiplex networks\cite{kim2013coevolution, lee2014threshold}. In the study of networks constructed from empirical data, the WTM has been used to examine phenomena such as protest recruitment\cite{gonzalez2011dynamics} and adoption of technology\cite{valente1996social}. There are also many variants of the WTM that caricature adoption behavior in different ways; these variants include thresholds that rely on the total number of neighbors \cite{centola2007complex}, a multi-stage threshold model\cite{melnik2013multi}, on--off thresholds\cite{dodds2013limited}, threshold models with memory\cite{dodds2004universal, shrestha2014message, gleeson2016effects}, and a threshold model that incorporates node states (through ``synergistic'' effects) from nodes other than nearest neighbors \cite{juul2017}.

There are many possible reasons why somebody may wait before adopting an idea, buying a product, etc. Possibilities include unawarenss of an innovation, awareness but not yet deciding to adopt something, a ``decision'' to adopt something but laziness before changing behavior\cite{herbers1981time}, and so on. Different social-influence models have different adoption rules that codify behavioral latency (i.e., a delay before adopting a behavior) in different ways, and different adoption rules generate different patterns of growth of the fraction of adopters over time. The study of different patterns in an adoption process is an important aspect of research on dynamical processes on networks\cite{valente2015diffusion, gonzalez2011dynamics, borge2011structural}. For example, in the WTM, the time at which a node adopts an innovation depends both on the node's threshold and on the adoption status of its neighbors (and hence on network architecture), and this affects the temporal evolution of the fraction of adopters in a network. In the present paper, our goal is to incorporate response times into spreading models such as threshold models: even if a threshold is met or exceeded, there is often a delay until a behavior is adopted. This can be due to personal inertia or to other factors. In this paper, we introduce a \emph{timer} to model the tendency of individuals to wait some amount of time between deciding to adopt a behavior and actually adopting it, and we investigate how the incorporation of timers (especially ones that are heterogeneous in a population) changes qualitative dynamics of the WTM.

The rest of our paper is organized as follows. In Section \ref{sec:timer_model}, we generalize the WTM so that nodes have both an associated adoption threshold and an associated timer. We illustrate the effects of incorporating a timer on small networks in Section \ref{sec:timer_model_on_small_networks} and on large networks in Section \ref{sec:timer_model_on_large_scale}. In Section \ref{sec:analysis}, we use a pair approximation to do some analysis on a WTM with a timer. In Section \ref{sec:adoption_path_in_large_networks}, we examine ``adoption paths'' for this model on large networks. We conclude in Section \ref{sec:conclusion}. We include further discussions and calculations in appendices.


\section{Timer models} \label{sec:timer_model}

According to the Oxford English Dictionary, a ``timer'' is an automatic mechanism for activating an object at a preset time\cite{dictionary2008oxford}. We apply this concept to a discrete-stage social-influence model on a network by defining a ``timer model'' as a model in which a node adopts an innovation after a preset number of discrete time steps, where a countdown starts after some other condition (e.g., peer pressure on the node matching or exceeding some other threshold) has been met. In a timer model, each node in a network has an associated timer that is drawn from some probability distribution. Once the timer of a node is ``triggered'' (e.g., when the node meets some adoption condition), its counter starts counting down to $0$ the next time the node is updated, and it changes its state when its timer hits $0$. We use a discrete-time setting, so each timer starts decrementing one time unit after all other adoption conditions are satisfied. For example, if node $v_i$'s timer $\tau_{v_i} \in \mathbb{Z}_{\geq 0}$ is triggered at time $t=t'_{v_i}$, it changes its state at time $t = t'_{v_i}+\tau_{v_i}+1$, as that is one time unit after the timer countdown reaches $0$. This is the first time that node $v_i$ is considered for updating of its state after its ``timer-adoption condition'' (at time $t = t'_{v_i}+\tau_{v_i}$) occurs. 

If there is no adoption condition other than a timer-adoption condition, the timers of all nodes are triggered at time step $t=0$ (so they start decrementing at $t = 1$), and the adoption process terminates when the largest timer hits $0$. Therefore, the timer of a node is equal to the time of adoption of the node, and the cumulative distribution function of the timer distribution describes the adoption process; network structure plays no role in this case. Such a naive timer model already illustrates Everett Rogers's ideas about the spread of innovations \cite{rogers1958categorizing, rogers2010diffusion}, in which different adopter categories (innovators, early adopters, early majority, late majority, and laggards) are determined only by their different adoption times.

We are interested in incorporating the idea of timers (especially heterogeneous ones) into models of spreading, and in this paper we use the WTM for concreteness.


\subsection{WTM with timers} \label{sec:wtm_with_timer}

One can add a timing mechanism to any existing social-influence model in which nodes adopt an innovation based on the states of other nodes in a network. Let's consider what happens if we add a timing mechanism to the WTM. The WTM is a binary-state model \cite{watts2002simple,gleeson2013binary} in which a node can have state $s \in \{0, 1\}$. The WTM has monotonic dynamics, as a node's state can change from $0$ to $1$, but any node that attains state $1$ remains at that state for all time. When a node first changes its state from $0$ to $1$, we say that it ``adopts'' some behavior or idea. The adoption condition of a node in the WTM is that at least a fraction $\phi_{v_i}$ of its neighbors have previously adopted the behavior. The parameter $\phi_{v_i}$ is the ``threshold'' for node $v_i$, and the condition that at least this fraction of $v_i$'s neighbors have adopted is the threshold-adoption condition. In the WTM with timers, a node $v_i$ must meet its threshold-adoption condition and also its timer-adoption condition to adopt a behavior: the fraction of adopted neighbors of node $v_i$ must be at least its threshold $\phi_{v_i}$ and then its timer $\tau_{v_i}$ must hit $0$. 
Update rules for nodes can either be synchronous or asynchronous\cite{porter2016dynamical}; in our study, we use synchronous updating, in which all nodes are updated simultaneously during each discrete time step.


\subsection{Adoption paths} \label{sec:adoption_path}

\begin{figure}[h!]
	\centering
	\includegraphics[width=0.5\textwidth]{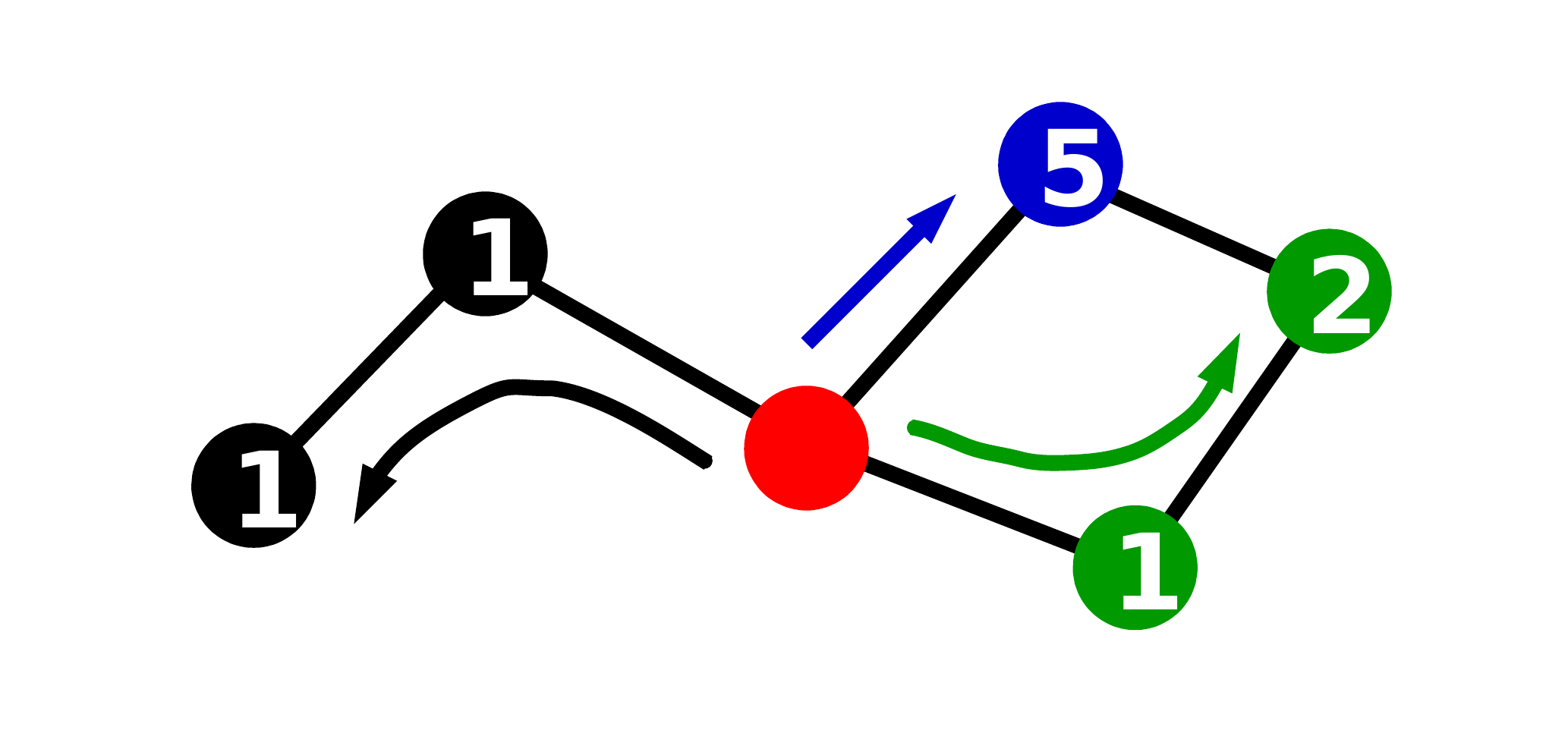}
	\caption{\raggedright Illustration of different adoption paths in threshold model with a timer. Each node $v_i$ is assigned a timer $\tau_{v_i}$ and a threshold $\phi_{v_i}$, where $i \in \{1, \cdots, 5\}$. We initiate node $v_0$ in the state $1$, so we do not need to assign it a threshold or a timer. Suppose that all nodes $v_i$ (with $i \in \{1, \cdots, 5\}$) have the same threshold $\phi=0.5$. An adoption of any node immediately triggers the timers of its neighbors.}
\label{fig:various_adoption_paths}
\end{figure}

We study the WTM with timers on undirected, unweighted networks; and we trace what we call \emph{adoption paths}, which are directed paths in a network through which an adoption is transmitted (see Fig.~\ref{fig:various_adoption_paths}). A directed path in a network is a sequence $(v_1, v_2, \dots , v_n)$ of nodes in which all nodes are distinct, and for $i \in \{1, \dots, n-1\}$, node $v_i$ is adjacent to $v_{i+1}$ via an edge from $v_i$. The length $l = n-1$ of a path is the number of edges that comprise the path. In an adoption path, node $v_i$ is adjacent to $v_{i+1}$ only if the adoption of $v_i$ triggers the timer $\tau_{v_{i+1}}$ of $v_{i+1}$. We call $v_1$ the ``root'' of the adoption path, as it initiates the spread of adoptions.\footnote{We use the term ``root'' in a different way from its conventional usage in graph theory.} The time $t'_{v_i}$ at which the timer of a node $v_i$ gets triggered is the sum of timers of all preceding nodes in the adoption path plus the number of synchronous time steps to trigger the timers of preceding nodes:
\begin{align}   \label{eqn:time_to_trigger_node_timer}
	t'_{v_i} = \sum_{j=1}^{i-1} \tau_{v_j} + i-1\,.
\end{align}
In Section \ref{sec:adoption_path_in_large_networks}, we use the unidirectional property of adoption paths to scrutinize how adoptions spread on large networks for the WTM with timers.

An adoption path in the WTM with timers terminates when
\begin{enumerate}
	\item[(i)]{the last node to adopt has degree $1$, or}
	\item[(ii)]{all neighbors of the last node to adopt have their timers triggered before the last node adopts.}
\end{enumerate}
In Fig.~\ref{fig:various_adoption_paths}, we show an example of different adoption paths that terminate after meeting either condition (i) or condition (ii). The dark brown node $v_0$ in the center is a seed node that has state $1$ at time $t=0$, and the other nodes are in state $0$ at $t=0$. Each node $v_i$ is assigned a timer $\tau_{v_i}$ and a threshold $\phi_{v_i}$, where $i \in \{1, \cdots, 5\}$. Node $v_0$ is the seed node, so it starts in the adopted state (and does not need a timer or threshold value). Nodes $v_i$ (for $i \in \{1, \cdots, 5\}$) have a threshold $\phi=0.5$, so an adoption of any node immediately triggers the timers of its neighbors. The number written above each node indicates its timer value. If we run the WTM with timers on this network, we obtain three adoption paths: $(v_0, v_1, v_2)$, $(v_0, v_3, v_4)$, and $(v_0, v_5)$. The lengths of these three adoption paths are $2$, $2$, and $1$, respectively. All adoption paths grow from the seed $v_0$, which is the root of all adoption paths. Each colored arrow represents the spread of an adoption through a distinct adoption path. The orange curved arrow terminates at time $T_{(v_0,v_1,v_2)}=4$; the last node to adopt has degree $k=1$, so it has no more neighboring nodes to influence (i.e., condition (i) is satisfied). The green arrow terminates at time $T_{(v_0, v_3, v_4)}=5$, and the blue arrow terminates at time $T_{(v_0,v_5)}=6$. Each of these adoption paths exhibits one of the two possible scenarios to satisfy condition (ii): for the blue adoption path, all neighbors of the last node to adopt are in state $1$; for the green adoption path, all neighbors of the last node to adopt are in state $0$, but the timers of all neighbors have already been triggered. The former scenario is also the condition for the termination of an adoption path in the original WTM, but the latter scenario is a novel feature of the WTM with timers. 


\section{WTM with timers on small networks} \label{sec:timer_model_on_small_networks}

Clearly, adding timers to the WTM in general will delay adoption processes. The presence of a homogeneous timer merely delays the adoption of each node for exactly the same number of time steps. Suppose that we run the WTM without timers on an arbitrary network, and it takes $T_\mathrm{WTM}$ time steps to reach a steady state, in which no more nodes can adopt. On the same network, if we run the WTM with homogeneous timers $\tau_\mathrm{hom}$, it now takes $T_\mathrm{WTM}(\tau_\mathrm{hom}+1)$ time steps to reach a steady state, as every node is delayed by the same amount of time. However, if timers are heterogeneous, different nodes have their adoptions delayed by different amounts of time, and it is less straightwaord to calculate the time to steady state in relation to $T_\mathrm{WTM}$.

\begin{figure}[h!]
	\centering
		\includegraphics[width=0.5\textwidth]{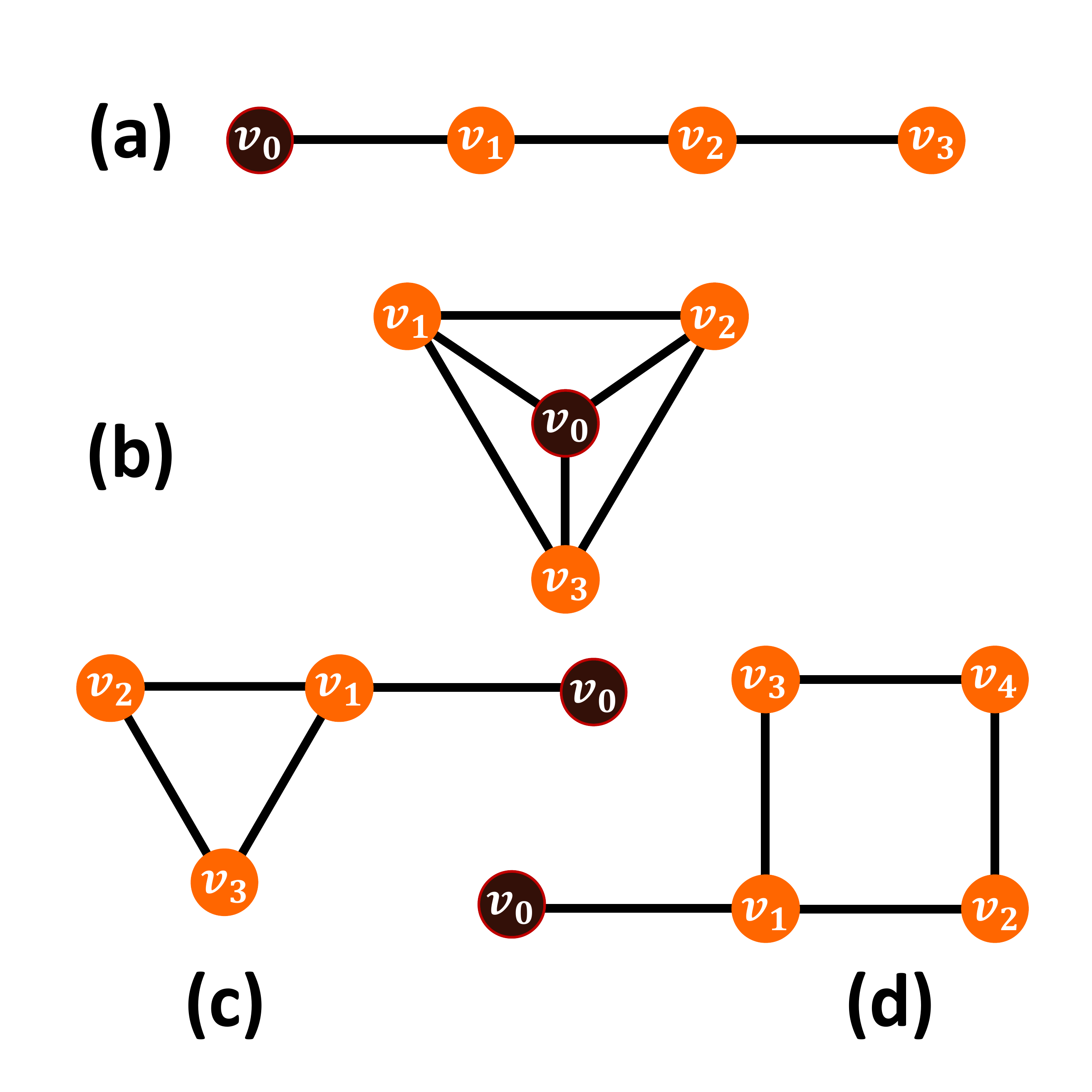}
	\caption{\raggedright Small examples to illustrate the effect of incorporating timers in the WTM. The dark brown node $v_0$ is the seed node with state $s=1$ at $t=0$, and the orange nodes are in state $s=0$ at $t=0$. The thresholds are homogeneous, with $\phi=0.1$ for each node.}
\label{fig:small_ex}
\end{figure}

We calculate the time $T_\mathrm{hom}$ to steady state for the WTM with homogeneous timers and the mean time $\langle T_\mathrm{het} \rangle$ to steady state for the WTM with heterogeneous timers in four small example networks in Fig.~\ref{fig:small_ex}. Suppose that all nodes have a homogeneous threshold of $\phi=0.1$, which is small enough so that any node in Fig.~\ref{fig:small_ex} adopts the state $s=1$ once even one of its neighbors is in the adopted state. Using a homogeneous threshold enables us to disentangle the effect of timers from the effect of a heterogeneous threshold. Because all nodes have the same positive threshold, we need a seed (node $v_0$ in Fig.~\ref{fig:small_ex}) in state $1$ at time $t=0$ to initiate the spread of adoptions.

\begin{table}[h!]
\centering
\caption{\label{tab:small_examples}
\raggedright Comparison between the time it takes for all nodes to adopt a behavior
(i.e., for a network to reach the fully-adopted state) 
for the small networks in Fig.~\ref{fig:small_ex} for the WTM with homogeneous timers $\tau_\mathrm{hom}=4$ and heterogeneous timers $\tau_\mathrm{het} \in \{2, 4, 6\}$ if the number of non-seed nodes (i.e., the orange nodes in Fig.~\ref{fig:small_ex}) is $3$, and $\tau_\mathrm{het} \in \{1, 3, 5, 7\}$ if the number of non-seed nodes is $4$. The mean value of the heterogeneous timers is $\langle \tau=4\rangle = \tau_\mathrm{hom}$. We calculate the mean time $\langle T_\mathrm{het} \rangle$ to steady state for heterogeneous timers by averaging over all possible configurations of timers with the given set of heterogeneous timers. 
In comparing homogeneous and heterogeneous timers in these networks, we also indicate if the adoption orders of nodes and/or adoption paths can change (i.e., if they can be different in the two scenarios) from what occurs in the WTM without timers.
}
\begin{tabular}{lcccc>{\centering\arraybackslash} m{1.5cm} >{\centering\arraybackslash} m{1.5cm}}
& $T_\mathrm{WTM}$\footnote{Time to steady state in the original WTM (i.e., without timers)} & $T_\mathrm{hom}$\footnote{Mean time to reach the fully-adopted state when the timers are homogeneous} & $\langle T_\mathrm{het} \rangle$\footnote{Mean time to reach the fully-adopted state averaged over all possible configurations of heterogeneous timers} & $\frac{\langle T_\mathrm{het} \rangle}{T_\mathrm{hom}}$ & Change of adoption order 
& Change of adoption paths 
\\
\hline
(a) & 3 & 15 & 15 & 1 & \xmark & \xmark \\
(b) & 1 & 5 & 7 & 1.4 & \vmark & \xmark \\
(c) & 2 & 10 & 11.33 & 1.133 & \vmark & \xmark \\
(d) & 3 & 15 & 13.66 & 0.91 & \vmark & \vmark \\
\end{tabular}
\end{table}

In Fig.~\ref{fig:small_ex}a, we show a one-dimensional (1D) lattice with a seed node at its left end, so adoption occurs from left to right. The time $T$ to steady state is
\begin{align} \label{eqn:ttss_a}
	T &= 3 + \tau_{v_1}+\tau_{v_2}+\tau_{v_3}\,,
\end{align}
where $\tau_{v_i}$ is the timer of node $v_i$. We need to add $3$ because there are $3$ nodes other than the seed, and it takes $1$ synchronous time step to trigger the timer of a node. In Table \ref{tab:small_examples}, we show results for this network when the nodes have a homogeneous timer with $\tau_\mathrm{hom} = 4$ and heterogeneous timers $\tau_\mathrm{het} \in \{2, 4, 6\}$ (note that the mean timer value is the same as in the homogeneous case), and we compare the time $T_\mathrm{hom}$ to steady state for the WTM with homogeneous timers and the mean time $\langle T_\mathrm{het} \rangle$ to steady state for all possible timer configurations of the WTM with heterogeneous timers. In this example, the time to steady state is simply a function of the sum of timers of all nodes because adoptions spread from each node to its neighbor on the right. Therefore, for a given mean timer value, the time to steady state is the same regardless of whether the timer values are distributed homogeneously or heterogeneously.
That is, the ratio $\langle T_\mathrm{het} \rangle/T_\mathrm{hom} = 1$. Note that neither the adoption order of the nodes nor the adoption path $(v_0, v_1, v_2, v_3)$ can change even if the timers are distributed heterogeneously in the 1D lattice. Adoption always starts from the left end, and it spreads to the right one node at a time, regardless of how the timers are distributed.

Using the same homogeneous threshold value $\phi=0.1$ as above, let's now consider what happens in a $4$-clique. In a $4$-clique, all nodes are adjacent to each other, so $v_1$, $v_2$, and $v_3$ are triggered simultaneously by the seed node $v_0$ at $t=0$. Therefore, with timers, the time $T$ to steady state is
\begin{align} \label{eqn:ttss_b}
	T &= 1 + \max (\tau_{v_1}, \tau_{v_2}, \tau_{v_3})\,.
\end{align}
We need to add $1$ because it takes $1$ synchronous time step to start a countdown after a timer is triggered. In a $4$-clique, the adoption paths $(v_0, v_1)$, $(v_0, v_2)$, and $(v_0, v_3)$ do not change for any assignment of timers. Moreover, the spread of adoptions in different adoption paths are independent of each other. Therefore, the adoption path that includes the node with the largest timer terminates last, determining the time to steady state. Hence, $T_\mathrm{het}$ is always larger than $T_\mathrm{hom}$ for a $4$-clique if the mean of the timer distribution is the same for the homogeneous and heterogeneous timers. 
In Table~\ref{tab:small_examples}, we show an example with the same homogeneous and heterogeneous timers as in Fig.~\ref{fig:small_ex}a, and we see that $\langle T_\mathrm{het} \rangle/T_\mathrm{hom} > 1$. We can generalize this to $k$-cliques with any value of $k$; the time to steady state in a $k$-clique is determined by the largest timer, so $\langle T_\mathrm{het} \rangle/T_\mathrm{hom} > 1$ if the mean of the timer distribution is kept constant.

The example in Fig.~\ref{fig:small_ex}c has a seed node adjacent to a square. The adoption of node $v_1$ triggers the timers of nodes $v_2$ and $v_3$ simultaneously, and whichever one adopts earlier triggers the timer of node $v_4$. The time $T$ to steady state is
\begin{widetext}
\begin{equation*} \label{eqn:ttss_d}
	T=\left\{
\begin{array}{l l}
  \tau_{v_1} + \max(2 + \tau_{v_3}, 3 + \tau_{v_2}+\tau_{v_4})\,, & \quad \text{if $\tau_{v_2} \leq \tau_{v_3}$\,,}\\
  \tau_{v_1} + \max(2 + \tau_{v_2}, 3 + \tau_{v_3}+\tau_{v_4})\,, & \quad \text{if $\tau_{v_2} > \tau_{v_3}$\,.}
	\end{array} \right.
\end{equation*}
\end{widetext}
We need to add $2$ if adoption paths $(v_0,v_1,v_3)$ or $(v_0,v_1,v_2)$ determine the time to steady state, and we need to add $3$ if adoption paths $(v_0,v_1,v_2,v_4)$ or $(v_0,v_1,v_3,v_4)$ determine the time to steady state. If the timers are homogeneous, the adoption of node $v_1$ triggers the timers of both nodes $v_2$ and $v_3$ simultaneously, and the simultaneous adoption of $v_2$ and $v_3$ triggers the timer of $v_4$, so the adoption paths are $(v_0,v_1,v_2,v_4)$ and $(v_0,v_1,v_3,v_4)$. However, if the timers are heterogeneous and the timers of $v_2$ and $v_3$ are different, there are changes in both the adoption order of the nodes and the adoption paths. Suppose, for example, that the timer of node $v_2$ is smaller than that of node $v_3$. In this case, node $v_2$ adopts before $v_3$, and it triggers the timer of $v_4$; therefore, the adoption paths are $(v_0,v_1,v_2,v_4)$ and $(v_0,v_1,v_3)$, and the one that takes longer time to terminate determines the time to steady state. In Table \ref{tab:small_examples}, we compare the time $T_\mathrm{hom}$ to steady state for a homogeneous timer $\tau_\mathrm{hom}=4$ and the mean time $\langle T_\mathrm{het} \rangle$ to steady state for heterogeneous timers $\tau_\mathrm{het} \in \{ 1, 3, 5, 7 \}$, and we see that $\langle T_\mathrm{het} \rangle < T_\mathrm{hom}$, in contrast to the other examples in Fig.~\ref{fig:small_ex}.

In fact, a network with a fixed seed node and fixed threshold assignments gives different adoption paths for different distributions of timers only if the network includes at least one cycle with $4$ or more nodes. Consider a node $v_i$ that is triggered by the adoption of node $v_j$ in adoption path $X$ for one timer distribution and by the adoption of node $v_k$ (with $v_k \neq v_j$) in adoption path $Y$ for another timer distribution. All adoption paths --- regardless of the timer distribution --- must share the same root (the seed node), because all adoptions spread from the seed. Therefore, adoption paths $X$ and $Y$ must share at least two nodes: $v_i$ and the seed node. Therefore, the network has a cycle that includes the seed node, $v_i$, and the other nodes in $X$ and $Y$. Therefore, a network that has different adoption paths for different distributions of timers must have a cycle of length at least $4$.

Because adoption paths can be different for heterogeneous timers than for a homogeneous timer, if the thresholds of all nodes in a network are sufficiently small so that any node adopts once at least one of its neighbors has adopted, an adoption path with nodes that have small timers tends to be long, and vice versa. Suppose that a node $v_i$ adopts if any one of its neighbors $v_j \in \Gamma(v_i)$ adopts, and suppose further that each neighbor belongs to a different adoption path of the same length. The time $t''_{v_i}$ at which node $v_i$'s timer is triggered is then
\begin{align}  \label{eqn:time_to_trigger_node_timer_multiple_neighbors}
  	t''_{v_i} = \min_{v_j \in \Gamma(v_i)}(t'_{v_j} + \tau_{v_j}) + 1 \,,
\end{align}
where $t'_{v_j}$ is the time that the timer of $v_j$ is triggered (see Eq.~(\ref{eqn:time_to_trigger_node_timer})). Node $v_i$ thereby becomes part of the adoption path that has the smallest sum of node timers, and this adoption path becomes longer than the other ones. Thus, in networks with cycles of length at least $4$, adoption paths with small mean timers tend to be long, and the adoption paths with large mean timers tend to be short. For example, in Fig.~\ref{fig:small_ex}c, the node with a smaller timer ($v_2$ in this case) is part of the longer adoption path, and the node with the larger timer ($v_3$ in this case) is part of the shorter adoption path. We will use the relationship between the length of an adoption path and the timer values of its nodes in Section \ref{sec:dissemination_tree} when we investigate long adoption paths with small mean timer values in large random networks. To compel nodes with small timers to be part of long adoption paths, it is important for their thresholds to be sufficiently small to adopt once at least one of their neighbors adopts. Otherwise, nodes with large timers can be part of longer adoption paths. See Appendix~\ref{app:large_threshold} for more details.


\section{Timers on large random networks} \label{sec:timer_model_on_large_scale}

We now incorporate both homogeneous and heterogeneous timers into a spreading process on large networks. Specifically, we examine the WTM with a timer on the largest connected component (LCC) of random networks with $N=10,000$ nodes (i.e., of ``size'' $10,000$). For many of our random networks, we use a homogeneous threshold $\phi$ such that a node whose degree is equal to the graph's mean degree needs only a single adopted neighbor to be triggered (as in the small networks in Fig.~\ref{fig:small_ex}). Because all nodes have the same threshold, we choose a seed adopted node uniformly at random at $t=0$ to initiate the spread of adoptions. In our simulations, we report sample means of many realizations to give an idea of ensemble expectations in the $N \rightarrow \infty$ limit.


\subsection{Timer distribution and dynamics of the WTM with a timer} \label{sec:timer_distribution}

\begin{figure*}[t!]
	\centering
	\includegraphics[width=\textwidth]{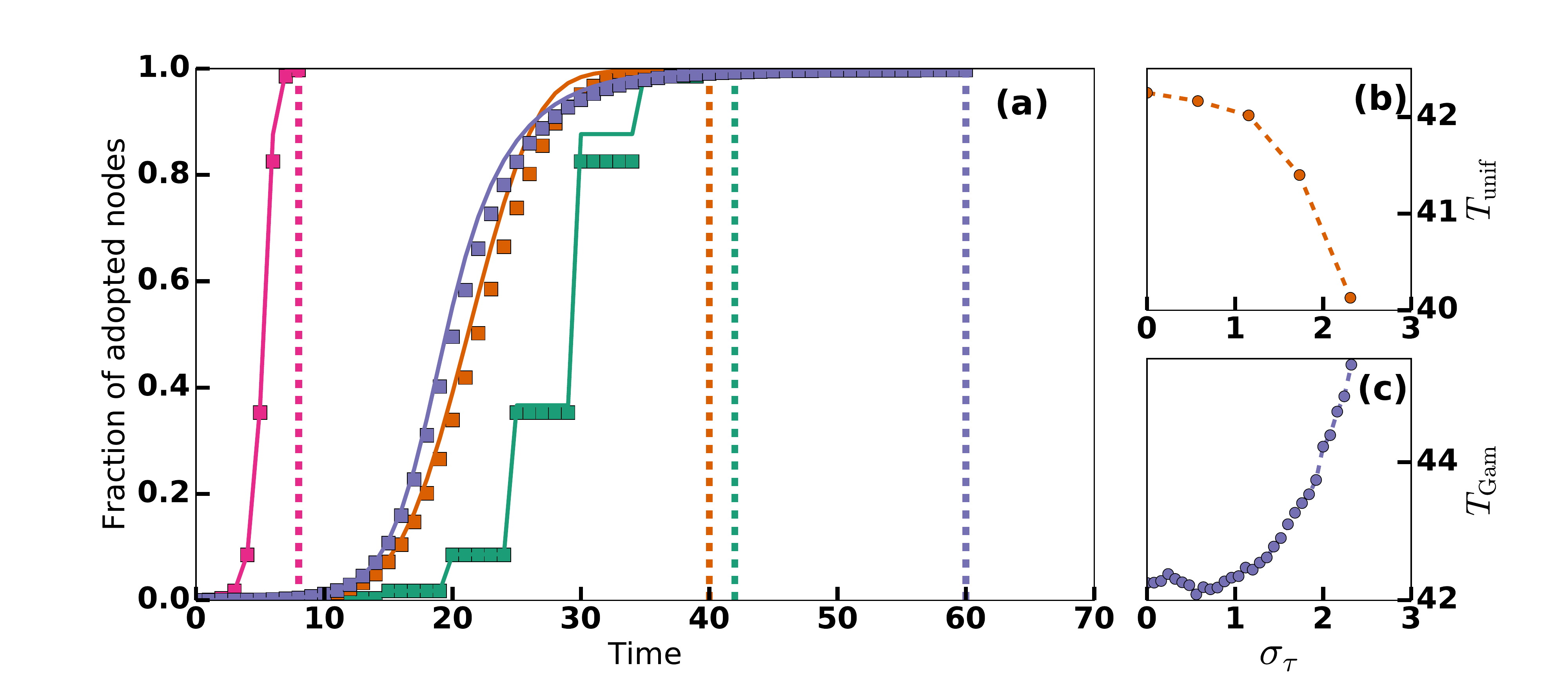}
	\caption{\raggedright (a) Adoption curves of the WTM with timers. The pink curve is without timers (i.e., the original WTM), the green curve is when the timers are homogeneous with $\tau=4$, the orange curve is with heterogeneous timers $\tau$ that are distributed uniformly at random from the set $\{0, 1, \dots, 8\}$, and the blue curve is for heterogeneous timers given by integers that we round down from a random variable that follows the Gamma distribution with mean $\mu_{\tau}=4$ and standard deviation $\sigma_{\tau}=4$. Squares are the results of numerical simulations, the dashed lines mark the times at which the adoption process of the corresponding color reaches a steady state, and the solid curves are results from an analytical approximation (see Section \ref{sec:analysis}). (b,c) Change of the times to steady state ($T_\mathrm{unif}$ and $T_\mathrm{Gam}$, respectively) by increasing the standard deviation $\sigma$ when timers are distributed (b) uniformly at random and (c) approximately according to a Gamma distribution. We obtain our numerical results by averaging over $1,000$ realizations of the WTM with timers on $G(N,p)$ ER networks with $N=10,000$ nodes and edge probability $p=0.0006$. To isolate the effects of incorporating timers, we use the same $1,000$ ER networks for each of the $4$ different cases.}
\label{fig:homo_vs_hetero_timers_on_ER}
\end{figure*}

We first consider the WTM with a timer on the LCC of $G(N,p)$ Erd\H{o}s--R\'{e}nyi (ER) networks with $N=10,000$ nodes and edge probability $p=0.0006$ (and thus an expected mean degree of $z=6$), and we compare the adoption curves --- the progress of the adopted fraction $\rho(t)$ of nodes at time $t$ --- between homogeneous and heterogeneous timers. In Fig.~\ref{fig:homo_vs_hetero_timers_on_ER}a, the pink curve shows the adoption process of the WTM without a timer, and the green curve shows the adoption process of the WTM with a homogeneous timer $\tau=4$. As the figure shows, at each point that adoption occurs, the green curve is simply delayed for $4$ time steps from the pink curve. The green adoption curve thus has a stair-like shape. If we employ asynchronous updating rather than synchronous updating (so that we choose some number of nodes uniformly at random at each time step to update their states \cite{porter2016dynamical}), the change in dynamics is not simply a delay for each adoption, as the randomness in node choice changes the adoption order \footnote{If we add a timer to any stochastic process on a network (e.g., with a non-deterministic update rule), a homogeneous timer does not in general simply delay adoption process, because the adoption order of adoption can change due to the randomness.}. See [\onlinecite{fennell2016}] for a discussion of discrete versus continuous dynamics in contagion models on networks.

The orange curve in Fig.~\ref{fig:homo_vs_hetero_timers_on_ER}a is the adoption curve of the WTM with heterogeneous timers on ER networks of size $N=10,000$ in which we select a timer $\tau \in \{0,1,\cdots,8\}$ uniformly at random. The mean $\mu_\tau$ of the timers is $4$, which is the same as for the homogeneous timer in the figure. The obvious difference from the heterogeneous timers is that the adoption curve is now much smoother (compare the orange and green curves). This arises amidst the change in adoption order from the heterogeneous timers; nodes that adopt simultaneously when timers are homogeneous now adopt at different times, and there are now fewer nodes that adopt simultaneously. (Note that some nodes that adopted at different times with a homogeneous timer now adopt simultaneously.) We also see that the time $T_\mathrm{unif}$ to steady state (orange dashed line) when the heterogeneous timers are distributed uniformly at random is earlier than the time $T_\mathrm{hom}$ to steady state (green dashed line) for homogeneous timers. In ER networks, we thus see that uniformly random heterogeneous timers accelerate the adoption process compared to the case of homogeneous timers. Increased heterogeneity in the distribution of the timers accelerates the adoption process even further. As one can observe from Fig.~\ref{fig:homo_vs_hetero_timers_on_ER}b, increasing the standard deviation $\sigma_\tau$ of the uniformly randomly distributed timers accelerates the time to steady state on ER networks.

Importantly, it is not true that heterogeneously-distributed timers necessarily accelerate an adoption process compared to homogeneous timers. For example, the blue curve in Fig.~\ref{fig:homo_vs_hetero_timers_on_ER} shows the adoption process of the WTM with timers given by integers that we determine by rounding down from numbers drawn uniformly at random from a Gamma distribution\cite{weisstein2004gamma} with the same mean $\mu_\tau=4$ as the uniformly randomly distributed timers and with standard deviation $\sigma_\tau=4$. The blue dashed line marks the time $T_\mathrm{Gam}$ at which the blue adoption curve reaches a steady state, and we observe that it is located to the right of the green dashed line $T_\mathrm{hom}$. In Fig.~\ref{fig:homo_vs_hetero_timers_on_ER}c, we show that the increase of standard deviation of the Gamma distribution decelerates the time to steady state on ER networks, in stark contrast to our observations in Fig. \ref{fig:homo_vs_hetero_timers_on_ER}b for timers that are distributed uniformly at random.

\begin{table}
\caption{\label{tab:adoption_percentage}
\raggedright Time to reach certain fractions of adopted nodes for the WTM on ER networks with homogeneous timers, heterogeneous timers distributed uniformly at random, and timers determined using a Gamma distribution and then rounded down to an integer.}
\centering
\begin{ruledtabular}
\begin{tabular}{c||c|cc|cc}
$\rho^*$\footnote{Fraction of adopted nodes} & $t_\mathrm{hom}$\footnote{Time to reach $\rho^*$ when timers are homogeneous.} & $t_\mathrm{unif}$\footnote{Time to reach $\rho^*$ when the timers are distributed uniformly at random.} & $\frac{t_\mathrm{unif}}{t_\mathrm{hom}}$ & $t_\mathrm{Gam}$\footnote{Time to reach $\rho^*$ when the timers are determined using a Gamma distribution and then rounded down to an integer.} & $\frac{t_\mathrm{Gam}}{t_\mathrm{hom}}$ \\
\hline
0.5 & 29.31 & 21.97 & 0.75 & 20.05 & 0.68 \\
0.6 & 29.52 & 23.18 & 0.79 & 21.21 & 0.72 \\
0.7 & 29.74 & 24.48 & 0.82 & 22.59 & 0.76 \\
0.8 & 29.95 & 25.97 & 0.87 & 24.43 & 0.82 \\
0.9 & 34.48 & 28.10 & 0.81 & 27.56 & 0.80 \\
$\rho_{\mathrm{st}}$\footnote{Fraction of adopted nodes at steady state.} & 42.41 & 40.21 & 0.95 & 60.44 & 1.43 \\
\end{tabular}
\end{ruledtabular}
\end{table}

Although the time to steady state for heterogeneous timers can become either larger or smaller than $T_\mathrm{hom}$, depending on the timer distribution, we observe that the majority of nodes adopt noticeably earlier when the timers are distributed either uniformly at random or using a Gamma distribution (and then rounded down to an integer)\footnote{Henceforth, we will usually state that these timers are determined using a Gamma distribution rather than always stating explicitly that we round down to an integer.} than they do for homogeneous timers. In Table \ref{tab:adoption_percentage}, we compare the time $t$ when the adopted fraction $\rho(t)$ reaches at least a certain fraction $\rho^*$ of nodes in networks when incorporating a homogeneous timer ($t_\mathrm{hom}$), heterogeneous timers distributed uniformly at random ($t_\mathrm{unif}$), and heterogeneous timers determined using a Gamma distribution ($t_\mathrm{Gam}$). We update nodes synchronously, so we do not in general have exactly the adoption fraction $\rho(t) = \rho^*$ at any time. In this case, we find the time $T'$ at which the fraction $\rho(T')$ of adopted nodes first exceeds $\rho^*$. We then estimate the time $t_{\rho=\rho^*}$ at which the fraction of adopted nodes reaches $\rho^*$ as
\begin{equation}
  \label{eqn:time_to_certain_fraction}
	t_{\rho=\rho^*} = T' - 1 + \frac{\rho^* - \rho(T'-1)}{\rho(T')-\rho(T'-1)}\,.
\end{equation}

As we show in Table \ref{tab:adoption_percentage}, the fraction $\rho(t)$ of adopted nodes tends to evolve faster when incorporating either uniformly randomly distributed timers or Gamma-distributed timers than for homogeneous timers, although Gamma-distributed timers take a longer time to reach the steady-state adoption fraction $\rho_{\mathrm{st}}$. This illustrates that a cascade --- the spread of adoptions from a small seed fraction of nodes to a much larger fraction of nodes --- can occur earlier when one considers heterogeneous timers than for homogeneous timers. In Section \ref{sec:adoption_path_in_large_networks}, we investigate adoption paths and give evidence for how incorporating heterogeneous timers in the WTM can make the majority of nodes adopt earlier than when considering homogeneous timers.
 

\section{Analysis} \label{sec:analysis}

We present an analytical approximation for the temporal evolution of the fraction of adopted nodes of the WTM with timers. To do this, we use a pair approximation, whose application to the WTM and its variants have been studied at length \cite{gleeson2007seed, gleeson2008cascades, hackett2013cascades, melnik2013multi,porter2016dynamical}. A pair approximation of the WTM was first developed by Gleeson and Cahalane \cite{gleeson2007seed}, who built on a method to study the zero-temperature random-field Ising model on Bethe lattices\cite{dhar1997zero}. Gleeson and Cahalane's pair approximation agrees well with the temporal evolution of the WTM, and it takes into account pairwise interaction betweens nodes \cite{gleeson2012accuracy}. We generalize their pair approximation to examine the temporal evolution of the adopted fraction of nodes in the WTM with timers.


\subsection{Pair approximation of the WTM}

We consider a pair approximation of the WTM for undirected, unweighted networks. We assume that our networks are locally tree-like, so that, asymptotically, cycles can be ignored (and there should be few short cycles in empirical networks) \cite{melnik2011}. Using these assumptions, it has been shown \cite{gleeson2007seed, gleeson2013binary} that one can approximate the evolution of the fraction $\rho(t)$ of adopted nodes in a network by calculating the probability that a node chosen uniformly at random has adopted at time $t$.

\begin{figure}[h!]
		\includegraphics[width=0.5\textwidth]{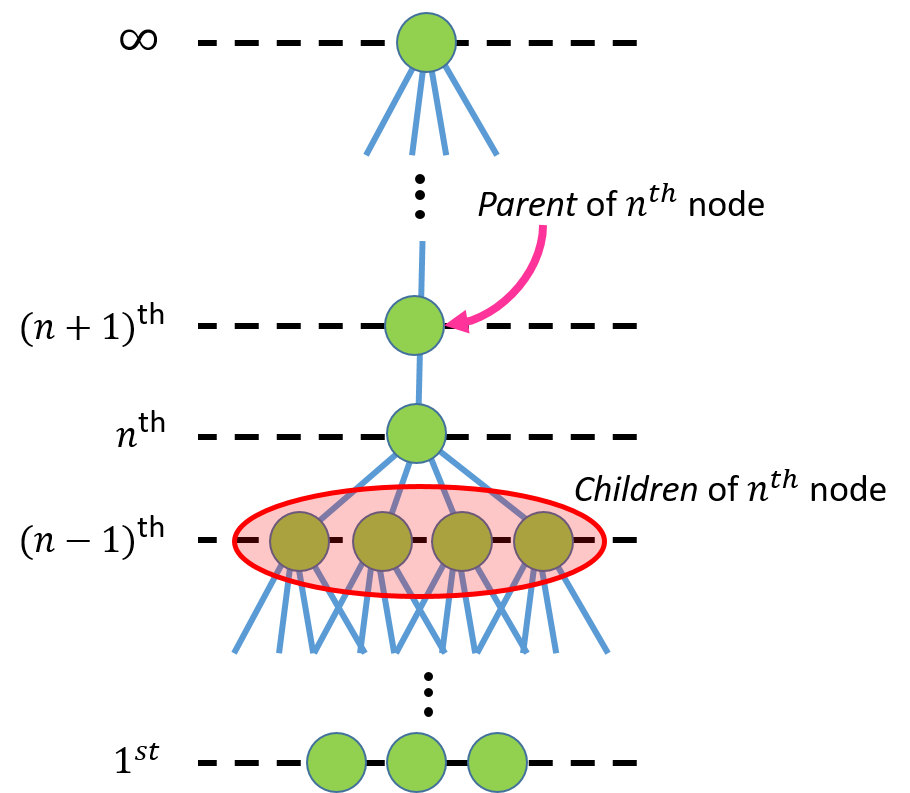}
	\caption{\raggedright An illustration of level-by-level spreading of adoption behavior in a network.
	}
\label{fig:analytics_diagram}
\end{figure}

To calculate the probability of a node to adopt, we first rearrange the network into a tree with the chosen node at the top level (i.e., level $\infty$) and its neighbors on the next level (see Fig.~\ref{fig:analytics_diagram}). In this way, the position of a node in the tree is determined by the distance between it and the top-level node. A node's ``parent'' is a neighbor located one level higher, and its ``children'' are its neighbors located one level lower. 

Because the threshold-adoption condition is that there are at least the threshold fraction of adopted neighbors, the probability of the top-level node to adopt is determined by the fraction of adopted children, whose probability of adoption is in turn determined by their childrens' adoption status, and so on. We thus write
\begin{align} \label{eqn:pair_approx_watts}
	\rho(t) &= \rho_0 + (1-\rho_0)\sum^{\infty}_{k=1}P_k\sum^{k}_{m=0}{k \choose m} \nonumber \\
	&\quad \times q_\infty(t)^m[1-q_\infty(t)]^{k-m}F(m, k)\,,
\end{align}
where $\rho_0$ is the fraction of seed nodes, $k$ is the degree, $P_k$ is the probability that the node degree is $k$ (i.e., $\{P_k\}$ is the degree distribution), $m$ is the number of adopted children, and $F(m,k)$ is the (neighborhood-influence) response function \cite{watts2007influentials, lopez2008social, hackett2013cascades}, which corresponds to the probability that a node satisfies the threshold-adoption condition for a given $m$ and $k$. As we discussed in Section \ref{sec:timer_model}, the threshold-adoption condition of a node $v_i$ is that the fraction ${m}/{k}$ of adopted neighbors is at least its threshold $\phi_i$. Therefore, for the WTM, the response function $F(m,k)$ is the probability for a node to have a threshold lower than ${m}/{k}$, which we can obtain from the cumulative distribution function of the thresholds. Finally, the term $q_n(t)$ is given by
\begin{align} \label{eqn:node_level_n_active_watts}
	q_{n}(t) &= \rho_0 + (1-\rho_0)\sum^{\infty}_{k=1}\frac{k}{z}P_k\sum^{k-1}_{m=0}{k-1 \choose m} \nonumber \\
&\times q_{n-1}(t)^m[1-q_{n-1}(t)]^{k-1-m}F(m, k)\,,
\end{align}
where $z$ is the mean degree of a network. Equation (\ref{eqn:node_level_n_active_watts}) gives the probability that a node at level $n$ of the tree has adopted at or before time $t$ (i.e., that it is in the adopted state at time $t$), conditional on its parent node at level $(n+1)$ being unadopted. The rationale behind the construction of $q_{n}(t)$ is as follows: A node $v_i$ chosen uniformly at random at level $n$ is in the adopted state if either
\begin{itemize}
  \item{the node is a seed, which occurs with probability $\rho_0$;}
  \item{or the node is not a seed, which occurs with probability $(1-\rho_0)$, but it meets the threshold-adoption condition at or before time $t$. (This condition is to have a threshold lower than ${m}/{k}$ when the node has $k-1$ child nodes at the $(n-1)^{th}$ level, with $m$ among them having adopted before time $t$.}
\end{itemize}
The factor $k-1$ comes from the fact the node has an unadopted parent at level $(n+1)$, so a node with degree $k$ and threshold ${m}/{k}$ should have $m$ adopted nodes among $k-1$ child nodes at level $n-1$; this yields the term ${k-1 \choose m}q_{n-1}(t)^m[1-q_{n-1}(t)]^{k-1-m}$. Finally, we need to sum over all possible $k$, because the nodes at level $n$ have various degrees that follow the degree distribution $\{P_k\}$. One reaches the nodes at level $(n-1)$ by following an edge between a child at level $(n-1)$ and the node at level $n$, so we use the excess degree distribution $\{{k}P_k/z\}$.

The above pair approximation shows good agreement with numerical simulations of the WTM\cite{gleeson2007seed, gleeson2013binary}, especially in large networks that are locally tree-like. This pair approximation has been generalized to several situations, including the study of the WTM on networks with community structure (including with heterogeneous communities) \cite{gleeson2008cascades, melnik2014dynamics} and other forms of clustering \cite{gleeson2009bond, hackett2013cascades}.


\subsection{Pair approximation of the WTM with timers}

We cannot simply use (\ref{eqn:pair_approx_watts}) and (\ref{eqn:node_level_n_active_watts}) for the WTM with timers, as a timer affects the time that a node adopts. Therefore, we need to understand the effect of timers in nodes' adoption and modify the pair-approximation equations accordingly. In the WTM with timers, a node with timer $\tau$ waits $\tau$ time steps after its threshold fraction of neighbors have adopted before it adopts. In other words, the adoption of a node is determined by both (i) its timer $\tau$ and (ii) the fraction of neighbors that have adopted $\tau$ time steps before the current time step. We need to consider both of these facets to derive a pair approximation for the WTM with timers.

The condition associated with (i) is determined by the response function $G(m,k,\tau)$ of the WTM with timers that decides whether a node has satisfied both the threshold-adoption condition and the timer-adoption condition. The response function $G(m,k,\tau)$ is the probability that a node has threshold less than ${m}/{k}$ and has timer $\tau$. It is given by
\begin{align} \label{eqn:timer_response_func}
	G(m, k, \tau) = F(m,k)[C_\tau(\tau)-C_\tau(\tau-1)]\,,
\end{align}
where $F(m,k)$ is the response function of the WTM and $C_\tau$ is the cumulative distribution function of the timers.

One can satisfy the condition associated with (ii) if a sufficient fraction ${m}/{k}$ of the children at level $(n-1)$ have adopted $\tau$ time steps before the current time step $t$. The probability that a child in level $n-1$ has adopted $\tau$ time steps ago is $q_n(t-\tau)$. Combining the conditions from (i) and (ii), we can express the adoption condition of a node with degree $k$, threshold ${m}/{k}$, and timer $\tau$ at time $t$ as
\begin{align} \label{eqn:prob_timer_adoption_condition}
	{k \choose m}q_n(t-\tau)^m[1-q_n(t-\tau)]^{k-m}G(m, k, \tau)]\,.
\end{align}

Similar to equations (\ref{eqn:pair_approx_watts}) and (\ref{eqn:node_level_n_active_watts}), we need to sum over all $m$, $k$, and $\tau$. We thereby obtain
\begin{widetext}
\begin{align}
	\rho(t) &= \rho_0 + \sum^t_{\tau=0}\left[(1-\rho_0)\sum^{\infty}_{k=1}\sum^{k}_{m=0}{k \choose m}q_\infty(t-\tau)^m[1-q_\infty(t-\tau)]^{k-m}G(m, k, \tau)\right], \label{eqn:pair_approx_timer} \\
	q_{n+1}(t) &= \rho_0 + \sum^t_{\tau=0}\left[(1-\rho_0)\sum^{\infty}_{k=1}\frac{k}{z}P_k\sum^{k-1}_{m=0}{k-1 \choose m}q_n(t-\tau)^m[1-q_n(t-\tau)]^{k-1-m}G(m, k, \tau)\right] \label{eqn:node_level_n_active_timer}\,.
\end{align}
\end{widetext}

As we saw in Fig.~\ref{fig:homo_vs_hetero_timers_on_ER}a, a numerical evaluation (solid curve) of the algebraic equations (\ref{eqn:pair_approx_watts},\ref{eqn:node_level_n_active_watts}) derived from our theory shows good agreement with direct numerical simulations (squares) of the WTM with timers for ER networks.

One can calculate a cascade condition\cite{watts2002simple} by linearizing equation (\ref{eqn:node_level_n_active_timer}). For a given mean threshold and mean degree in a network, a cascade condition determines whether there is a global cascade\cite{watts2002simple, gleeson2007seed}, in which a small seed fraction $\rho_0$ of adopted nodes results in a large value of $\rho_\infty = \lim_{t\rightarrow\infty}\rho(t)$. Because the WTM dynamics are monotonic, if a node adopts in the limit $t\rightarrow \infty$, the node also adopts in the WTM with a timer in the limit $t\rightarrow \infty$. Consequently, in the $t\rightarrow \infty$ limit, $q_n(t)$ of the WTM with timers and $q_n(t)$ of the WTM yields the same value $q_n(\infty)$. Therefore, the cascade condition of the WTM with timers is identical to the cascade condition of the WTM without timers.


\section{Adoption paths in large networks} \label{sec:adoption_path_in_large_networks}

In Section \ref{sec:timer_model}, we argued that the WTM with a homogeneous timer $\tau_\mathrm{hom}$ has exactly the same adoption paths as the adoption paths of the WTM without timers, because homogeneous timers do not change the adoption order of nodes but instead merely delays adoption times uniformly by $\tau_\mathrm{hom}$. Therefore, the time $T_\mathrm{hom}$ to achieve a steady state is delayed to $T_\mathrm{hom}=T_\mathrm{WTM}(1+ \tau_\mathrm{hom})$, where $T_\mathrm{WTM}$ is the time to steady state when there are no timers. However, if the timers are distributed heterogeneously, the adoption order of nodes can change, so adoption paths can also change (depending on the network structure). Furthermore, the relationship between the time $T_\mathrm{het}$ to steady state for heterogeneous timers and $T_\mathrm{WTM}$ is more complicated than the relationship between $T_\mathrm{hom}$ and $T_\mathrm{WTM}$. In Section \ref{sec:timer_distribution}, we observed from simulations on ER networks that the WTM with timers distributed uniformly at random and timers determined using a Gamma distribution yield earlier adoptions for the majority of nodes adopt earlier than is the case for a homogeneous timer with the same mean (see Table \ref{tab:adoption_percentage}). In this section, we explore this issue in depth by investigating adoption paths of the WTM with timers for both synthetic and real-world networks.


\subsection{Stems and branches} \label{sec:stem_and_branch}

As we discussed in Section \ref{sec:timer_model}, an adoption path is a sequence of directed edges in which each edge indicates the flow of adoption from a node at one end to a node at the other end. All adoption paths grow from the seed, which we recall is the node that initiates the spread of adoptions in a network. Therefore, the seed is the root of all adoption paths. Among the adoption paths, which have various lengths, we pick a longest adoption path at steady state. We do not include the root as part of the adoption path, which we call a \emph{stem} of adoption spreading (or simply a ``stem''). Nodes in a stem are called \emph{stem nodes}. If there are two or more adoption paths that both have the largest length, we consider all of them to be stems. For the other adoption paths, we exclude the stem nodes; their remaining nodes are \emph{branches} of adoption spreading (or simply ``branches''). The main difference between stems and branches is that stems grow from a seed node, whereas branches grow from stems.

\begin{figure}
		\includegraphics[width=0.5\textwidth]{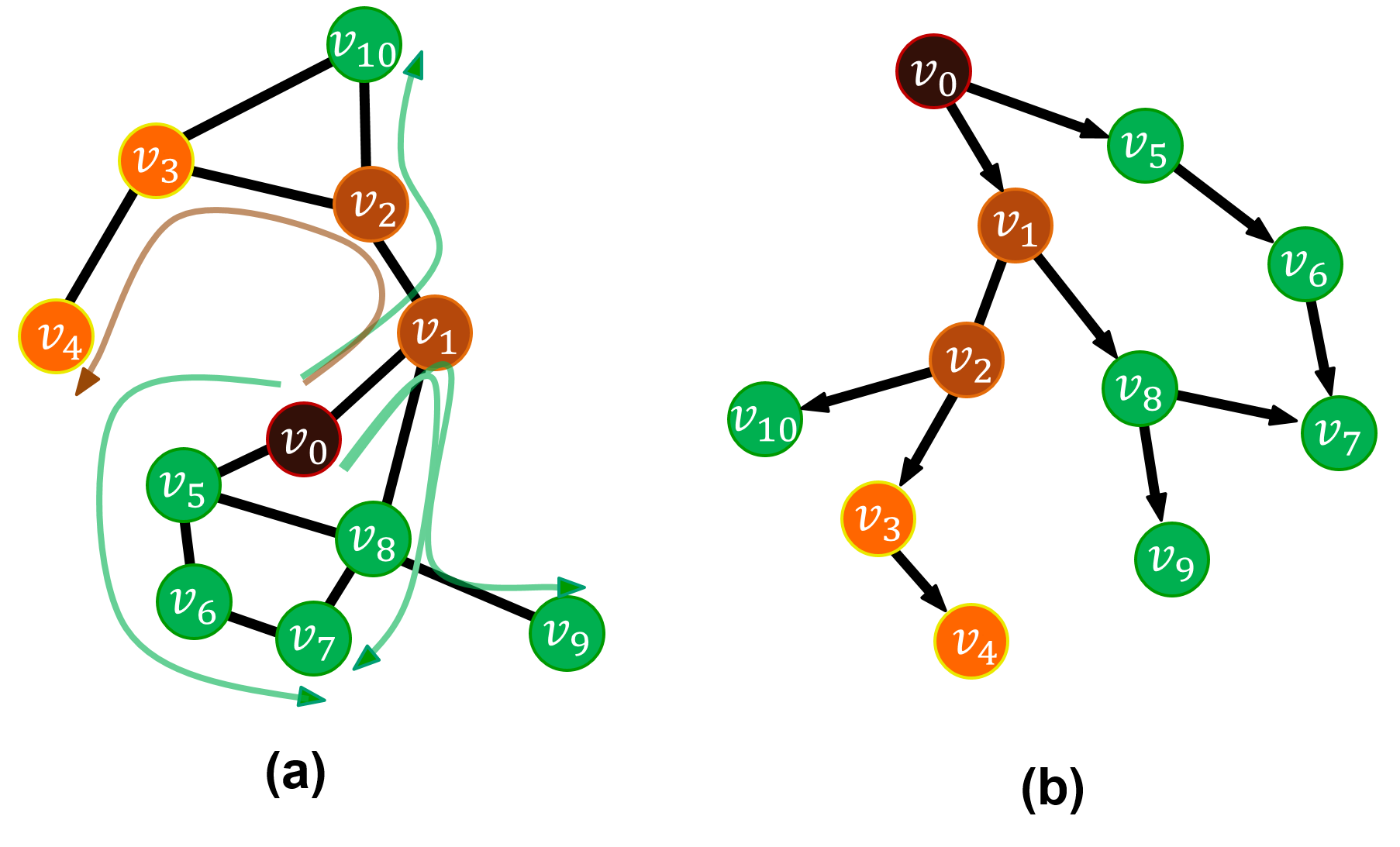}
		\caption{\raggedright (a) An example network on which we run the WTM with timers. Curved arrows illustrate the spread of adoptions, and each curved arrow represents an adoption path. The brown curved arrow is a stem, and the green curved arrows are branches. (b) Graphical illustration of a dissemination tree, based on the spread of adoptions in panel (a). Node $v_0$ is a seed, which is the root of the stem $(v_1, v_2, v_3, v_4)$. Nodes $v_0$, $v_1$, and $v_2$ in the stem initiate the adoption in branches; they are the roots of the branches $(v_5, v_6, v_7)$, $(v_8, v_7)$, $(v_8, v_9)$, and $(v_{10})$.}
\label{fig:illustration_stem_branch}
\end{figure}

We give an example of a stem and its branches in Fig.~\ref{fig:illustration_stem_branch}. Suppose that we run the WTM with timers on the network in Fig.~\ref{fig:illustration_stem_branch}a with node $v_0$ as a seed and that we obtain five adoption paths: $(v_0, v_1, v_2, v_3, v_4)$, $(v_0, v_5, v_6, v_7)$, $(v_0, v_1, v_8, v_7)$, $(v_0, v_1, v_8, v_9)$, and $(v_0, v_1, v_2, v_{10})$. Among the adoption paths, $(v_0, v_1, v_2, v_3, v_4)$ has the largest length, so the stem is $(v_1, v_2, v_3, v_4)$. The branches are $(v_5, v_6, v_7)$, $(v_8, v_7)$, $(v_8, v_9)$ and $(v_{10})$. Note that $v_7$ and $v_8$ each appear in two different adoption paths, because the adoption of $v_8$ triggers the timers of $v_7$ and $v_9$, and the timer of $v_7$ is simultaneously triggered by $v_6$ and $v_8$.

With the adoption paths, we can construct a tree (see Fig.~\ref{fig:illustration_stem_branch}b), which demonstrates how adoptions spread in the original network in Fig.~\ref{fig:illustration_stem_branch}a. We use the term \emph{dissemination tree} for a network composed of adoption paths.\footnote{Dissemination trees have been studied previously, though using other terminology\cite{gomez2010inferring,myers2010convexity,snowsill2011refining,ugander2012}. Reference [\protect\onlinecite{goel2015structural}] used the term ``diffusion tree'', but we use the word ``dissemination'' instead of ``diffusion'' to distinguish diffusion from other types of spreading dynamics (e.g., non-conservative contagion dynamics).} Because an adoption path is directed, a dissemination tree does not have a path back to the root and is thus a directed acyclic graph (DAG)\cite{newman2009networks}. However, as one can see in Fig.~\ref{fig:illustration_stem_branch}b, the underlying undrected graph of a dissemination tree can have a cycle of length at least $4$; in Fig.~\ref{fig:illustration_stem_branch}b, this cycle is $(v_0, v_1, v_8, v_7, v_6, v_5, v_0)$. However, a dissemination tree's underlying undirected graph cannot have triangular clustering (i.e., 3-node cycles). See Appendix~\ref{app:cycle} for further discussion. We investigate the characteristics of dissemination trees in Section \ref{sec:dissemination_tree}.


\subsection{Dissemination trees for synthetic networks} \label{sec:dissemination_tree}

\begin{figure*}[t!]
		\includegraphics[width=\textwidth]{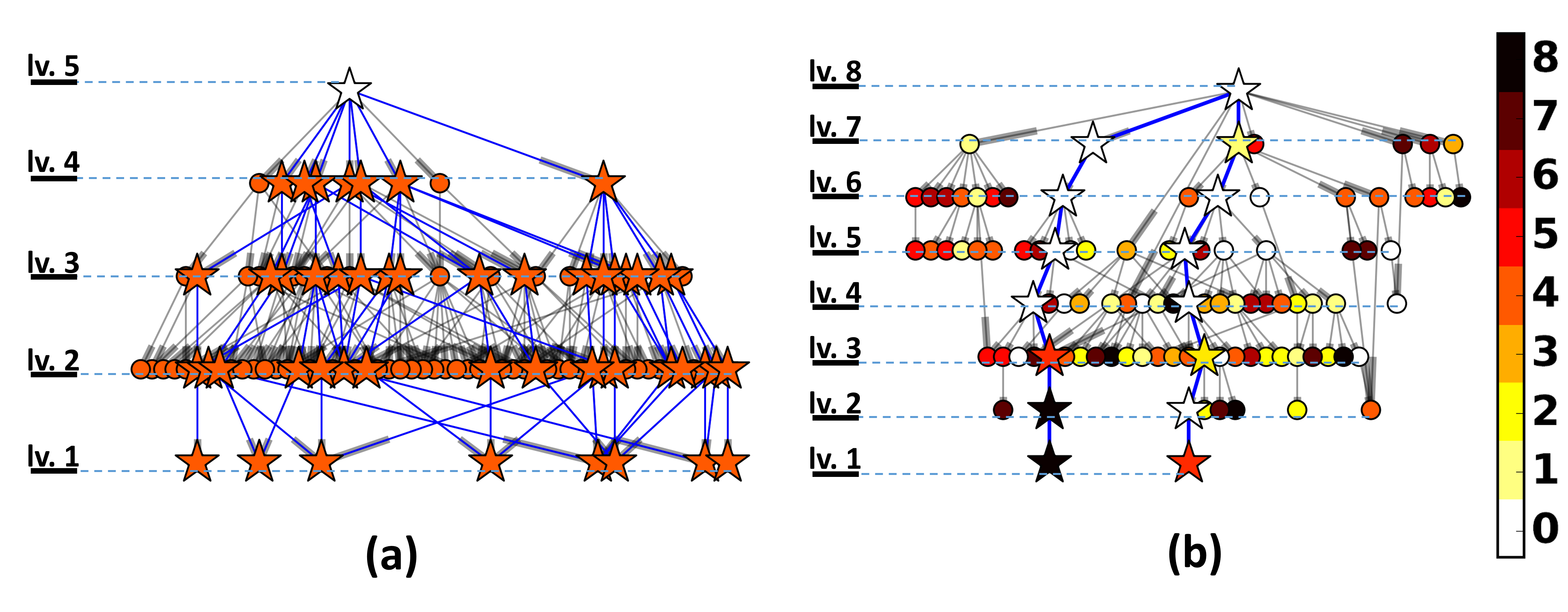}
	\caption{\raggedright Graphical representation of a dissemination tree for the WTM with (a) homogeneous timers and (b) heterogeneous timers. The edges are directed from higher levels to lower levels. We color edges in blue if they are part of a stem and in gray if they are part of a branch. The solid blue lines indicate edges that are part of stems, and the solid gray lines indicate edges that are part of branches. We color the nodes according to their time values, which range from $0$ (white) to $8$ (black). We use stars for nodes in stems and disks for nodes in branches. The number of levels in the dissemination tree with homogeneous timers is $5$, and the number of levels with heterogeneous timers is $8$.
	}
\label{fig:dissemination_tree}
\end{figure*}

In Fig.~\ref{fig:dissemination_tree}, we show graphical representations of dissemination trees that we obtain from running the WTM with timers on an ER network of size $N=100$, mean degree $z=6$, and a homogeneous threshold of $\phi=0.1$ for each node. We give examples with both homogeneous and heterogeneous timers, and we examine adoption paths and combine them to create a dissemination tree in each case. We place the root of the stem (i.e., the seed node) at the top of the tree, and we place the rest of the nodes in the subsequent levels based on their distance from the root. In Fig.~\ref{fig:dissemination_tree}a, we show a dissemination tree when timers are distributed homogeneously with $\tau=4$. In Fig.~\ref{fig:dissemination_tree}b, we show a dissemination tree when the timers are distributed heterogeneously (and, in particular, drawn uniformly at random from $\{0,1,\dots,8\}$). We color the nodes based on their timer values, which ranges from $0$ (white) to $8$ (black). We use stars for nodes in stems and disks for nodes in branches. We color the edges of stems in blue, and we use gray for edges in branches. When the timers are homogeneous, there are more stems (many different ones with the same maximal length), whose length is smaller than in the example with heterogeneous timers. When the timers are heterogeneous, we also observe that the timers of nodes in a stem tend to be small (lighter colors). As we discussed in Section~\ref{sec:adoption_path}, the timer of a node in an adoption path gets triggered instantaneously by the adoption of its neighbor that is adjacent via an in-edge (i.e., by the previous node in the adoption path). Thus, the time to terminate an adoption path is determined by the sum of timers of nodes in the adoption path [see Eq.~(\ref{eqn:time_to_trigger_node_timer})]. Consequently, if the mean of the node timers in an adoption path is small, an adoption tends to be transmitted through the path in a short amount of time. The mean timer value of the nodes in an adoption path thus helps determine how fast an adoption spreads in an adoption path. Therefore, the mean timer value of stem nodes being smaller than the mean value of branch-node timers suggests that adoption should tend to spread at a faster rate along stems than along the branches. In Table~\ref{tab:adoption_path_random_networks}, we show results for several families of random networks, and we observe qualitatively similar phenomona as in Fig.~\ref{fig:dissemination_tree}.

\begin{table*}
\caption{\label{tab:adoption_path_random_networks}
\raggedright Comparison between the the characteristics of stems and branches from running the WTM with timers on configuration-model networks (``Config''), generalized configuration-model networks with $3$-cliques (``Congen-$3$''), and generalized configuration-model networks with $4$-cliques (``Congen-$4$''). For all networks in this table, we start with configuration-model networks with $N=10,000$ nodes and degrees drawn from a Poisson distribution $P_k = {z^k e^{-z}}/{k!}$ with mean $z=6$. All nodes have a homogeneous threshold of $\phi=0.1$, so that a node with the mean degree adopts once one of its neighbors adopts. For the generalized configuration models, we consider different values of the edge--clique ratios $\alpha$ and $\beta$, where $\alpha$ determines the edge--clique ratio for a node of degree $k \geq 3$ in Congen-$3$ and $\beta$ determines the edge--clique ratio for a node of degree $k \geq 4$ in Congen-$4$. The notation $x_\mathrm{hom}$ indicates that we calculate the quantity $x$ for homogeneous timers, the notation $x_\mathrm{unif}$ indicates that we calculate the quantity $x$ for timers that are distributed uniformly at random, and the notation $x_\mathrm{Gam}$ indicates that we calculate the quantity $x$ when the timers are determined using a Gamma distribution. The quantity $T$ is the time to steady state, and $t_{\rho^*}$ is the time that it takes for adopted fraction $\rho^*$ to reach a certain fraction of adopted nodes. (In this table, we use $\rho^*=0.9$.) The quantity $\mu_{l}$ is the mean adoption path length, $\sigma_{l}$ is the standard deviation of the lengths of adoption paths, $v$ is the \emph{structural virality} (which is defined as the mean shortest path length in a dissemination tree), $a$ is the mean number of adoption paths, $f_{a_s}$ is the mean percentage of stems among adoption paths, $f_{a_b}$ is the mean percentage of branches among adoption paths, $r$ is the mean number of adopted nodes, $f_{r_s}$ is the mean percentage of stem nodes, $f_{r_b}$ is the mean percentage of branch nodes, $l_s$ is the mean stem length, and $l_b$ is the mean branch length. To compare the effects of different timer distributions, we augment the WTM with the different timer distributions on the same network with the same seed node and the same adoption-threshold distribution for each realization. Each reported value is a mean over $100$ simulations of the WTM with timers on networks generated independently for each simulation. 
}
\begin{ruledtabular}
\begin{tabular}{p{3cm}|||ccccccc|cccccc||}

\multicolumn{13}{c}{Homogeneous timer} \\
& $T_\mathrm{hom}$ & $t_\mathrm{0.9, hom}$ & $r_\mathrm{hom}$ & $a_\mathrm{hom}$ & $\mu_{l,\mathrm{hom}}$ & $\sigma_{l,\mathrm{hom}}$ & $v_\mathrm{hom}$
& $f_{r_s}$ ($\%$) & $f_{r_b}$ ($\%$) & $f_{a_s} ($\%$)$ & $f_{a_b} ($\%$)$ & $l_{s, \mathrm{hom}}$ & $l_{b, \mathrm{hom}}$ \\ \hline
Config & 42.76 & 32.97 & 9999.43 & 17505.69 & 7.72 & 0.36 & 7.15 & 4.44 & 95.56 & 1.07 & 98.93 & 9.55 & 4.13 \\
Congen-$3$ ($\alpha = 0.5$) & 44.06 & 34.30 & 9998.99 & 16754.55 & 7.91 & 0.37 & 7.31 & 4.70 & 95.30 & 1.25 & 98.75 & 9.81 & 4.30 \\
Congen-$3$ ($\alpha = 1.0$) & 46.32 & 35.43 & 9993.95 & 15770.20 & 8.16 & 0.37 & 7.54 & 3.02 & 96.98 & 0.62 & 99.38 & 10.26 & 4.57 \\
Congen-$4$ ($\beta = 0.5$) & 45.41 & 35.03 & 9987.32 & 17316.51 & 8.05 & 0.41 & 7.48 & 3.46 & 96.54 & 0.88 & 99.12 & 10.08 & 4.56 \\
Congen-$4$ ($\beta = 1.0$) & 52.17 & 38.78 & 9996.23 & 17250.30 & 8.72 & 0.43 & 8.17 & 1.66 & 98.34 & 0.31 & 99.69 & 11.43 & 5.57 \\
\hline
\multicolumn{13}{c}{Uniformly random distribution of timers} \\
& $T_\mathrm{unif}$ & $t_\mathrm{0.9, unif}$ & $r_\mathrm{unif}$ & $a_\mathrm{unif}$ & $\mu_{l,\mathrm{unif}}$ & $\sigma_{l,\mathrm{unif}}$ & $v_\mathrm{unif}$
& $f_{r_s}$ ($\%$) & $f_{r_b}$ ($\%$) & $f_{a_s} ($\%$)$ & $f_{a_b} ($\%$)$ & $l_{s, \mathrm{unif}}$ & $l_{b, \mathrm{unif}}$ \\ \hline
Config & 40.39 & 27.26 & 9999.43 & 7935.59 & 9.02 & 0.46 & 10.52 & 0.34 & 99.66 & 0.05 & 99.95 & 15.90 & 6.57 \\
Congen-$3$ ($\alpha = 0.5$)& 41.76 & 28.34 & 9998.99 & 7763.81 & 9.19 & 0.49 & 10.79 & 0.35 & 99.65 & 0.35 & 99.65 & 16.13 & 6.75 \\
Congen-$3$ ($\alpha = 1.0$)& 44.10 & 29.77 & 9993.95 & 7511.16 & 9.38 & 0.47 & 11.16 & 0.35 & 99.65 & 0.06 & 99.94 & 16.53 & 6.92 \\
Congen-$4$ ($\beta = 0.5$) & 42.81 & 27.13 & 9987.32 & 7798.33 & 9.29 & 0.54 & 11.03 & 0.35 & 99.65 & 0.06 & 99.94 & 16.30 & 6.84 \\
Congen-$4$ ($\beta = 1.0$) & 49.60 & 32.81 & 9996.23 & 7500.79 & 9.87 & 0.54 & 12.05 & 0.34 & 99.66 & 0.05 & 99.95 & 17.35 & 7.36 \\
\hline
\multicolumn{13}{c}{Gamma distribution (and then rounding down to an integer) of timers} \\
& $T_\mathrm{Gam}$ & $t_\mathrm{0.9, Gam}$ & $r_\mathrm{Gam}$ & $a_\mathrm{Gam}$ & $\mu_{l,\mathrm{Gam}}$ & $\sigma_{l,\mathrm{Gam}}$ & $v_\mathrm{Gam}$
& $f_{r_s}$ ($\%$) & $f_{r_b}$ ($\%$) & $f_{a_s} ($\%$)$ & $f_{a_b} ($\%$)$ & $l_{s, \mathrm{Gam}}$ & $l_{b, \mathrm{Gam}}$ \\ \hline
Config & 61.19 & 29.89 & 9999.43 & 7786.92 & 8.51 & 0.44 & 10.45 & 0.50 & 99.50 & 0.09 & 99.91 & 13.15 & 6.13 \\
Congen-$3$ ($\alpha = 0.5$)& 61.92 & 30.93 & 9998.99 & 7630.97 & 8.72 & 0.47 & 10.75 & 0.48 & 99.52 & 0.08 & 99.92 & 13.52 & 6.32 \\
Congen-$3$ ($\alpha = 1.0$)& 63.67 & 32.23 & 9993.95 & 7400.31 & 8.98 & 0.47 & 11.14 & 0.44 & 99.56 & 0.07 & 99.93 & 14.09 & 6.60 \\
Congen-$4$ ($\beta = 0.5$) & 62.69 & 31.75 & 9987.32 & 7701.32 & 8.88 & 0.52 & 11.00 & 0.45 & 99.55 & 0.08 & 99.92 & 13.85 & 6.46 \\
Congen-$4$ ($\beta = 1.0$) & 66.99 & 35.23 & 9996.23 & 7493.46 & 9.60 & 0.53 & 12.08 & 0.41 & 99.59 & 0.07 & 99.93 & 15.26 & 7.12 \\
\end{tabular}
\end{ruledtabular}
\end{table*}

In Table \ref{tab:adoption_path_random_networks}, we give a comparison between the characteristics of the dissemination trees for the WTM with a homogeneous timer $\tau_\mathrm{hom}=4$ versus the dissemination trees for the WTM with heterogeneous timers on different random-graph models. For the WTM with heterogeneous timers, we first consider timers drawn uniformly at random from $\{0,1,\dots,8\}$, and we then consider timers drawn from a Gamma distribution with mean $\mu_{\tau}=4$ and standard deviation $\sigma_{\tau}=4$ and then rounded down to an integer. We show results for $13$ diagnostics in Table \ref{tab:adoption_path_random_networks}: the time $T$ to steady state, the time $t_\mathrm{0.9}$ that it takes for at least $90\%$ of the nodes to adopt, the mean $r$ number of adopted nodes in the networks, the mean $a$ number of adoption paths, the mean $\mu_{l}$ and the standard deviation $\sigma_{l}$ of adoption-path lengths, \emph{structural virality}\cite{goel2015structural} $v$ (which is defined as the mean shortest-path length in a dissemination tree\footnote{Structural virality is one quantity that has been used to try to forecast the eventual size of a cascade, so it useful for studies of information dissemination and social influence. See \cite{nikhil2016eval} for a comparison of different ways --- including ones that are based on cascades --- to measure social influence in a network.}), and several stem-specific and branch-specific quantities. These latter quantities are the mean percentage $f_{a_s}$ of stems and the mean percentage $f_{a_b}$ of branches among adoption paths, the mean percentage $f_{r_s}$ of stem nodes and the mean percentage $f_{r_b}$ of branch nodes among adopted nodes, and the mean stem length $l_s$ and mean branch length $l_b$.

The network models that we use in Table \ref{tab:adoption_path_random_networks} are a configuration model \cite{fosdick2016configuring} and generalized configuration models that include cliques\cite{gleeson2009bond, hackett2013cascades}. We construct configuration-model networks by specifying a degree distribution $\{P_k\}$ and then connecting stubs (i.e., ends of edges) uniformly at random. To construct networks using a generalized configuration model, we embellish the above configuration model by incorporating cliques\cite{gleeson2009bond, hackett2013cascades}. We start with a configuration-model network with degree distribution $\{P_k\}$, but there is also a joint distribution $\gamma(k,c)$ that specifies the probability that a node chosen uniformly at random has degree $k$ and is in a clique of $c$ nodes (i.e., a $c$-clique). Note that $\gamma(k,c)=0$ for $k<c-1$, as a node with degree $k$ can only be a member of a $c$-clique if its degree is large enough to link to all $c-1$ neighbors in the clique.

For all networks in Table \ref{tab:adoption_path_random_networks}, we use configuration-model networks with $N=10,000$ nodes and degrees drawn from the Poisson distribution $P_k = {z^k e^{-z}}/{k!}$ with mean $z=6$. In the networks, each node has a homogeneous adoption threshold of $\phi=0.1$, so that a node with the mean degree adopts once one of its neighbors adopts. We use ``Config'' to denote a standard configuration model; ``Congen-$3$'' to denote a generalized configuration model with $3$-cliques and with joint distribution $\gamma(k,c) = \left[ (1-\alpha)\delta_{c,1} + \alpha\delta_{c,3} \right]P_k$ for $k \ge 3$ (where the parameter $\alpha$ determines the edge--clique ratio for a node of degree $k \geq 3$); and ``Congen-$4$'' to denote a generalized configuration model with $4$-cliques and with joint distribution $\gamma(k,c) = \left[ (1-\beta)\delta_{c,1} + \beta\delta_{c,4} \right]P_k$ for $k \ge 4$ (where $\beta$ determines the edge--clique ratio for a node of degree $k \geq 4$). 

As we discussed in Section~\ref{sec:timer_model_on_large_scale}, the time $T$ to steady state for the WTM with heterogeneous timers can be either shorter (for timers distributed uniformly at random) or longer (for Gamma-distributed timers) than for a homogeneous timer, but both choices of heterogeneous timer distributions have a smaller value than with a homogeneous timer for the time $t_{0.9}$ for at least $90\%$ of the nodes to adopt. Additionally, for both distributions of heterogeneous timers, we observe a smaller mean number $n$ of adoption paths than for a homogeneous timer. This, in turn, results in a shorter mean adoption path length $\mu_{l}$ for homogeneous timers than for heterogeneous ones, because for a fixed network size $N$, having a larger number of adoption paths leads to shorter mean adoption path lengths. Note that the number $n$ of adoption paths of a network can become larger than the number $r$ of adopted nodes in a network, because a node can occur in different adoption paths if the node triggers the timers of multiple nodes at the same time.

If timers are homogeneous, adoptions spread at the same rate for every adoption path, and adoption paths that terminate at the same time have the same lengths. However, for heterogeneous timers, adoption paths that terminate at the same time can have different lengths, because adoptions can spread at different rates. Therefore, the lengths of adoption paths can be more diverse when using heterogeneous timers than for homogeneous ones. As we see in Table~\ref{tab:adoption_path_random_networks}, the standard deviation $\sigma_{l}$ of adoption path lengths is larger for the WTM with heterogeneous timers than with homogeneous timers. That is, $\sigma_{l,\mathrm{unif}} > \sigma_{l,\mathrm{hom}}$, and $\sigma_{l,\mathrm{Gam}} > \sigma_{l,\mathrm{hom}}$ in our simulations.

Goel et al.\cite{goel2015structural} studied how viral content that spreads from peer to peer in large networks has a different spreading pattern from content that does not go viral, and they introduced the idea of structural virality to try quantify the virality of content from its spreading pattern. They defined structural virality $v$ as the mean shortest-path length in a dissemination tree, and a larger $v$ indicates that the mean distance between any two nodes chosen uniformly at random from a dissemination tree is larger. If a meme goes viral, it is reasonable that the mean distance between randomly-chosen nodes should be larger than for memes that do not become viral. We show structural viralities of dissemination trees in Table \ref{tab:adoption_path_random_networks}, and the results provide evidence that $v$ is larger for the WTM with heterogeneous timers than with homogeneous ones.

For the WTM with each of the distributions of timers that we examine, the stem-specific quantities $f_{r_s}$ and $f_{a_s}$ are smaller than the branch-specific quantities $f_{r_b}$ and $f_{a_b}$. In particular, for the two types of heterogeneous timers, both $f_{r_s}$ and $f_{a_s}$ in Table \ref{tab:adoption_path_random_networks} are less than $1$, and they are smaller than the corresponding quantities when we use a homogeneous timer. For a homogeneous timer, all adoption paths grow at the same rate, so the adoption path (or paths, if there is a tie) that grows for the longest time is the longest adoption path at steady state. Therefore, in this situation, all adoption paths that grow until steady state are stems. However, adoption paths grow at different rates for heterogeneous timers. Therefore, even if a stem terminates when a simulation reaches a steady state, not all adoption paths that grow until a steady state need to be stems. (A stem can terminate before a branch if the sum of timers of the stem nodes is smaller than the sum of branch-node timers.) We observe in our simulations that the mean percentage $f_{a_s}$ of stems is smaller for heterogeneous timers than for a homogeneous timer. We speculate that for most network structures the number of stems should be smaller for the WTM with heterogeneous timers than for the WTM with a homogeneous timer.

\begin{table*}
\centering
\caption{\label{tab:stem_branch_timer_profile}
\raggedright Comparison between the mean timer of stems and branches for the WTM model with a homogeneous timer, uniformly-randomly-distributed timers, and timers determined using a Gamma distribution. The networks that we use are a configuration model (Config), a generalized configuration model with $3$-cliques (Congen-$3$), and a generalized configuration model with $4$-cliques (Congen-$4$). For all networks, we use configuration-model networks with $N=10,000$ nodes and degrees drawn from a Poisson distribution $P_k = {z^k e^{-z}}/{k!}$ with mean $z=6$. Additionally, each node has a homogeneous threshold of $\phi=0.1$, so that a node with the mean degree adopts once one of its neighbors adopts. For the generalized configuration models, we consider different values of the edge--clique ratios $\alpha$ and $\beta$, where $\alpha$ determines the edge--clique ratio for a node of degree $k \geq 3$ in Congen-$3$ and $\beta$ determines the edge--clique ratio for a node of degree $k \geq 4$ in Congen-$4$. The notation $x_\mathrm{hom}$ indicates that we calculate the quantity $x$ for homogeneous timers, the notation $x_\mathrm{unif}$ indicates that we calculate the quantity $x$ for timers that are distributed uniformly at random, and the notation $x_\mathrm{Gam}$ indicates that we calculate the quantity $x$ for timers determined using a Gamma distribution. The quantity $\tau_s$ is the mean stem-node timer, $\tau_b$ is the mean branch-node timer, and $\langle \tau \rangle$ is the mean of all timers.
}
\begin{ruledtabular}
\begin{tabular}{p{2.75cm}||cc|cc|cc}

 & \multicolumn{2}{c}{Homogeneous timer} & \multicolumn{2}{c}{Uniformly random timers} & \multicolumn{2}{c}{Gamma-distributed timers} \\

 & $\frac{\tau_{s, \mathrm{hom}}}{\langle \tau \rangle}$ & $\frac{\tau_{b, \mathrm{hom}}}{\langle \tau \rangle}$ & $\frac{\tau_{s, \mathrm{unif}}}{\langle \tau \rangle}$ & $\frac{\tau_{b, \mathrm{unif}}}{\langle \tau \rangle}$ & $\frac{\tau_{s, \mathrm{Gam}}}{\langle \tau \rangle}$ & $\frac{\tau_{b, \mathrm{Gam}}}{\langle \tau \rangle}$ \\ \hline

Config & 1.0 & 1.0 & 0.29 & 1.00 & 0.42 & 1.01 \\
Congen-$3$ ($\alpha=0.5$) & 1.0 & 1.0 & 0.30 & 1.00 & 0.40 & 1.01 \\
Congen-$3$ ($\alpha=1.0$) & 1.0 & 1.0 & 0.33 & 1.00 & 0.40 & 1.00 \\
Congen-$4$ ($\beta=0.5$)  & 1.0 & 1.0 & 0.32 & 1.00 & 0.39 & 1.00 \\
Congen-$4$ ($\beta=1.0$) & 1.0 & 1.0 & 0.35 & 1.00 & 0.44 & 1.00 \\
\end{tabular}
\end{ruledtabular}
\end{table*}

To better understand the difference between the stems and branches of the networks in Table~\ref{tab:adoption_path_random_networks}, we investigate (1) the ratio ${\tau_{s}}/{\langle \tau \rangle}$ of the mean $\tau_{s}$ of timer values of nodes in stems to the mean $\langle \tau \rangle$ of timer values of all nodes, and (2) the ratio ${\tau_{b}}/{\langle \tau \rangle}$ of the mean $\tau_{b}$ of timer values of nodes in branches to $\langle \tau \rangle$ in Table \ref{tab:stem_branch_timer_profile}. For the WTM with a homogeneous timer, both $\frac{\tau_{s,\mathrm{hom}}}{\langle \tau \rangle}  = 1$ and $\frac{\tau_{b,\mathrm{hom}}}{\langle \tau \rangle} = 1$, so stems and branches grow at the same rate. When the timers are distributed uniformly at random, stems tend to grow at faster rates than the mean rate because $\frac{\tau_{s, \mathrm{unif}}}{\langle \tau \rangle} < 1$, whereas branches grow at slower rates than the mean rate because $\frac{\tau_{b,\mathrm{unif}}}{\langle \tau \rangle}$ is (slightly) larger than $1$. Similarly, when the timers are determined using a Gamma distribution, $\frac{\tau_{s, \mathrm{Gam}}}{\langle \tau \rangle} < 1$ and $\frac{\tau_{b,\mathrm{Gam}}}{\langle \tau \rangle}$ is slightly larger than $1$. The reason that there is only slight difference between $\tau_b$ and $\langle \tau \rangle$ for the heterogeneous timers that we study is that the percentages $r_b$ of branch nodes for heterogeneous timers are larger than $99\%$ of all nodes in networks, as we see in Table~\ref{tab:adoption_path_random_networks}.

To better understand why the mean $\tau_b$ is larger than $\langle \tau \rangle$ and $\tau_s$ is smaller than $\langle \tau \rangle$, it is useful to revisit our discussion from the end of Section \ref{sec:timer_model_on_small_networks} about the relationship between the mean timer size for nodes in an adoption path and the length of the adoption path. We know that a network can have different adoption paths for different timer distributions. Consider a scenario in which any node in a given network adopts if even one of its neighbors adopts. The nodes with small timers tend to be part of long adoption paths, and the nodes with large timers tend to be part of short adoption paths. From Table~\ref{tab:adoption_path_random_networks}, we know that the adoption paths for a homogeneous and heterogeneous timers are different for different distributions of timers, because the number $a$ of adoption paths entails different values for different timer distributions. The nodes in the networks in Table~\ref{tab:adoption_path_random_networks} have a homogeneous adoption threshold of $\phi=0.1$, so nodes with degree of $10$ or less will adopt once a single neighbor adopts. For example, in the configuration-model networks, the expected mean of the degree distribution is $z=6$, so there are a larger number of nodes with degree less than or equal to $10$ than with degree larger than $10$, and most of the nodes in these networks adopt if any of their neighbors adopt.  Accordingly, long adoption paths of dissemination trees in Table~\ref{tab:adoption_path_random_networks} are more likely to consist of nodes with small timers. Thus, the ratio ${\tau_{s}}/{\langle \tau \rangle}$ of the mean timer value of stem nodes to the mean timer value $\langle \tau \rangle$ of all nodes is smaller than $1$ for the WTM with heterogeneous timers, as we can see in Table~\ref{tab:stem_branch_timer_profile}.
  
\begin{figure}[h!]
	\centering
	\includegraphics[width=0.5\textwidth]{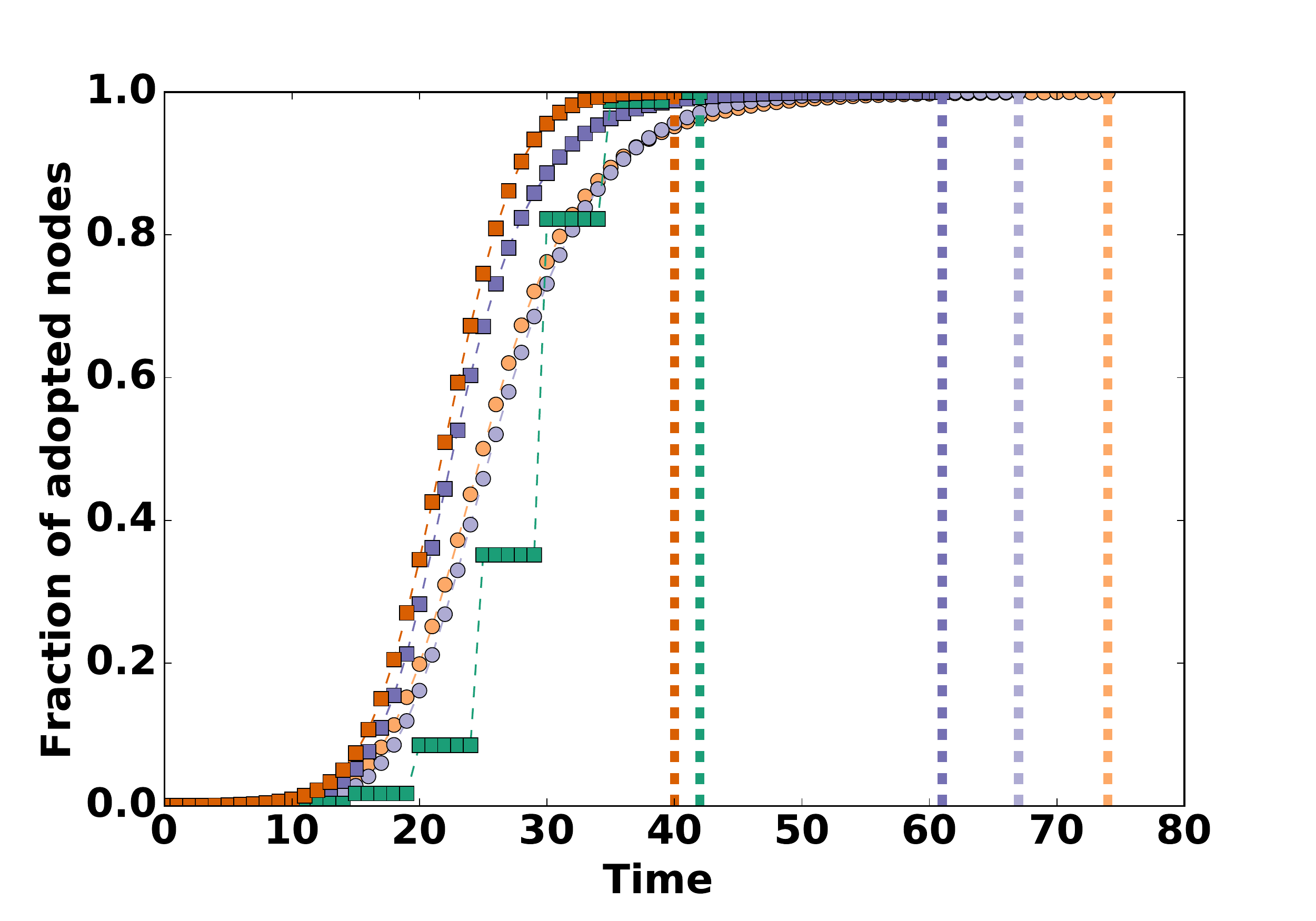}
	\caption{\raggedright Adoption curves of the WTM with timers and their adoption curves after we change the timers of stem nodes of the dissemination trees. The green curve is for the WTM with homogeneous timers with $\tau=4$, the orange curves are for the WTM with heterogeneous timers $\tau$ that are distributed uniformly at random from the set $\{0, 1, \dots, 8\}$, and the blue curves are for heterogeneous timers determined from the Gamma distribution with mean $\mu_{\tau}=4$ and standard deviation $\sigma_{\tau}=4$ and then rounding down to an integer. To isolate the effects of different distributions of timers, in each case, we run the WTM using the same networks with the same seed nodes. The dark orange and blue curves are before we change the timers in dissemination trees, and the corresponding light-colored curves are after we change those timer values (see the main text). The squares (for the dark curves) and disks (for the light curves) are the results of numerical simulations, and the dashed lines mark the times at which the adoption process of the corresponding color reaches a steady state. We obtain our numerical results by averaging over $1,000$ realizations of the WTM with timers on configuration-model networks with $N=10,000$ nodes and a Poisson degree distribution with mean $z=6$. For each realization, we generate an independent network. We also determine the seed node (uniformly at random) and timer values separately for each realization.
	}
\label{fig:slower_stem_on_config}
\end{figure}

Recall that (1) rates of adoption spreading in stems are smaller for the WTM with heterogeneous timer distributions than with a homogeneous timer distribution and (2) the rates of adoption spreading in branches are only slightly larger for heterogeneous timer distributions than for a homogeneous timer distribution. Therefore, in our calculations, it seems that stems play a significant role in spreading adoptions to the majority of nodes faster for heterogeneous timers than for a homogeneous timer. To support our speculation, we conduct the following experiment on the dissemination trees in Tables~\ref{tab:adoption_path_random_networks} and \ref{tab:stem_branch_timer_profile}. We know that each node in a dissemination tree has its own timer (i.e., the timer that is determined before the simulation starts) and it adopts at a certain time. Suppose that we change the timers of stem nodes of the dissemination tree to the mean value $\langle \tau \rangle$ of the timers (which is $\mu_{\tau}=4$ in our example). The times that the stem nodes' neighbors in the dissemination tree adopt then also change, as the time of adoption of a node in an adoption path is determined by the sum of its timer and the time of adoption of its predecessor node. (Note that we are not rerunning the WTM dynamics; instead, we are adjusting an adoption curve after a simulation.) In Fig.~\ref{fig:slower_stem_on_config}, we investigate how the adoption curves change because of the change of timers of stem nodes. The orange and the blue curves are both adoption curves for the WTM with heterogeneous timers on configuration-model networks with a Poisson degree distribution with mean $z=6$. For the orange curve, the timers are distributed uniformly at random from the set $\{0, 1, \dots, 8\}$. For the blue curve, the timers are determined from a Gamma distribution with mean $\mu_{\tau}=4$ and standard deviation $\sigma_{\tau}=4$ that we then round down to an integer. The curves with lighter colors are the adoption curves after we change the timers of the stem nodes of dissemination trees to the mean timer value. The light orange curve is for timers distributed uniformly at random, the light blue curve is for timers determined from a Gamma distribution, and the green curve (which we include to facilitate our comparison) is for the WTM with a homogeneous timer. We observe that intersections between the light-colored curves and the green curve occurs earlier than the corresponding ones between the dark-colored curves and the green curve. We also show in Table~\ref{tab:slower_stem} how the time to reach certain fractions $\rho^*$ of nodes calculated using Eq.~(\ref{eqn:time_to_certain_fraction}) changes by adjusting the timers of stem nodes. From both Fig.~\ref{fig:slower_stem_on_config} and Table~\ref{tab:slower_stem}, we observe that the adoption processes for the WTM with heterogeneous timers is delayed by changing the timers of stem nodes, even though they constitute fewer than $1\%$ of all nodes in the networks. In Appendix~\ref{app:slower_stem_on_other_random_networks}, we show our results for our synthetic networks from Table~\ref{tab:adoption_path_random_networks}. Although we have examined the above situation for specific families of networks, we believe that the idea that we just tested is relevant much more broadly. We test our model on real-world networks in Section \ref{sec:real_application}.

\begin{table}
\caption{\label{tab:slower_stem}
\raggedright Time to reach certain fractions of adopted nodes for the WTM with timers on configuration-model networks and the changes of the time to reach certain fractions of adopted nodes after we adjust the timers of stem nodes of dissemination trees for the WTM with timers. We consider configuration-model networks with $N=10,000$ nodes and a Poisson-degree distribution with mean $z=6$. All nodes have a homogeneous adoption threshold of $\phi=0.1$. For timer distributions, we use homogeneous timers with $\tau=4$, heterogeneous timers distributed uniformly at random from the set $\{0, 1, \dots, 8\}$, and timers determined using a Gamma distribution with mean $\mu_{\tau}=4$ and standard deviation $\sigma_{\tau}=4$ and then rounding down to an integer. To isolate the effects of different distributions of timers, in each case, we run the WTM using the same networks with the same seed nodes. We average over $1,000$ realizations, and we determine the seed node (uniformly at random) and timer values separately for each realization. We also generate an independent network for each realization.}
\centering
\begin{ruledtabular}
\begin{tabular}{c||c|cc|cc}
$\rho^*$\footnote{Fraction of adopted nodes} & $t_\mathrm{hom}$\footnote{Time to reach $\rho^*$ for the WTM with a homogeneous timer.} & $t_\mathrm{unif,WTM}$\footnote{Time to reach $\rho^*$ for the WTM with timers distributed uniformly at random.} & $t_\mathrm{unif,dis}$\footnote{Time to reach $\rho^*$ after we change the timers of stem nodes for timers that are distributed uniformly at random.} & $t_\mathrm{Gam,WTM}$\footnote{Time to reach $\rho^*$ for the WTM with timers determined using a Gamma distribution.} & $t_\mathrm{Gam,dis}$\footnote{Time to reach $\rho^*$ after we change the timers of stem nodes for a timers determined using a Gamma distribution.} \\
\hline
0.5 & 29.02 & 22.01 & 25.17 & 22.76 & 25.70 \\
0.6 & 29.82 & 23.10 & 26.71 & 23.89 & 27.24 \\
0.7 & 30.42 & 24.24 & 28.48 & 25.21 & 29.04 \\
0.8 & 31.26 & 25.54 & 30.79 & 26.97 & 31.36 \\
0.9 & 32.97 & 27.26 & 34.59 & 29.88 & 35.01 \\
$\rho_{st}$\footnote{Fraction of adopted nodes at steady state.} & 42.76 & 40.40 & 74.55 & 61.19 & 67.55 \\
\end{tabular}
\end{ruledtabular}
\end{table}


\subsection{Dissemination trees for real-world networks} \label{sec:real_application}

We now investigate dissemination trees for the WTM with timers on five {\sc Facebook100} networks \cite{traud2012social,traud2011}. Each network represents a Facebook friendship network at one university in the United States. In each of these networks, a node is an individual, and an unweighted, undirected edge represents a Facebook friendship between a pair of individuals. 
As one can see in Table~\ref{tab:adoption_path_real_networks}, these networks have larger mean degrees than those of the random networks from Table~\ref{tab:adoption_path_random_networks}. Therefore, to ensure that a node with degree equal to the mean degree adopts if a single one of its neighbors has adopted, we consider a homogeneous adoption threshold of $\phi=0.01$.
For each of the five networks, we observe qualitatively similar results as in the random-graph ensembles, although the {\sc Facebook100} networks have different structural characteristics (e.g., in terms of local clustering and community structure) than the random-graph models. We show our results in Table \ref{tab:adoption_path_real_networks}. We find that the percentage $f_{a_s}$ of stems is smaller than the percentage $f_{a_b}$ of branches, and that the numbers $n$ of adoption paths are larger for the WTM with homogeneous timers than with heterogeneous timers. Accordingly, for heterogeneous timers, the lengths $l_{s}$ of the stems and the lengths $l_{b}$ of the branches are both larger than for homogeneous timers. Additionally, $\frac{\tau_{s,\mathrm{het}}}{\langle \tau \rangle}<1$ and $\frac{\tau_{b,\mathrm{het}}}{\langle \tau \rangle}>1$.

As we can see from the paragraph above, stems tend to grow faster and become longer than branches for the WTM with heterogeneous timers. In other words, the characteristics of the stems and the branches for the five {\sc Facebook100} networks seem to be similar to those of the random networks that we studied previously. When timers are heterogeneous, stems grow faster and become longer, initiating adoption spreading in branches earlier than when the timers are homogeneous. Consequently, a majority of nodes adopt earlier for the WTM model with heterogeneous timers than with homogeneous ones. In Table \ref{tab:adoption_path_real_networks}, we show values for how long it takes for at least 90\% of the nodes to adopt. We observe analogous results for other values $\rho^*$ of adopted fractions; see Appendix~\ref{app:real_net_adoption_percentage} for a table of values $\rho^*$ ranging from $0.1$ to $1.0$.

\begin{table*}
\caption{\label{tab:adoption_path_real_networks}
\raggedright Comparison between stem and branch characteristics for the WTM model with timers on {\sc Facebook100} networks. 
The quantity $N$ is the number of nodes in a network, $z$ is the network's mean degree, and $\langle l \rangle$ is the mean shortest-path length between pairs of nodes. Each nodes has a homogeneous adoption threshold of $\phi=0.01$. The notation $x_\mathrm{hom}$ indicates that we calculate the quantity $x$ for homogeneous timers, the notation $x_\mathrm{unif}$ indicates that we calculate the quantity $x$ for timers that are are distributed uniformly at random, and the notation $x_\mathrm{Gam}$ indicates that we calculate the quantity $x$ for timers determined using a Gamma distribution (and then round down to an integer). The quantity $T$ is the time it takes to reach a steady state, and $t_{\rho^*}$ is the time that it takes for a fraction $\rho^*$ of nodes to adopt. (He use $\rho^*=0.9$ in this table; see see Appendix~\ref{app:real_net_adoption_percentage} for other values of $\rho^*$.) The quantity $a$ is the mean number of adoption paths, $\mu_{l}$ is the mean adoption path length, $\sigma_{l}$ is the standard deviation of the adoption path lengths, and $v$ is the structural virality. The quantity $f_{a_s}$ is the mean percentage of stems among adoption paths, $f_{a_b}$ is the mean percentage of branches among adoption paths, $l_s$ is the mean stem length, $l_b$ is the mean branch length, $\tau_s$ is the mean timer value of stem nodes, $\tau_b$ is the mean timer of branch nodes, and $\langle \tau \rangle$ is the mean value of all timers. Each quantity is a mean over $100$ realizations with different timers and seed nodes (which are determined uniformly at random).
 }
\begin{ruledtabular}
\begin{tabular}{p{2cm}|||ccc||cccccc|cccccc||}
\hline
& & & & \multicolumn{12}{c}{Homogeneous timer} \\


 & N & z & $\langle l \rangle$
 & $T_\mathrm{hom}$ & $t_\mathrm{0.9, hom}$ & $a_\mathrm{hom}$ & $\mu_{l,\mathrm{hom}}$ & $\sigma_{l,\mathrm{hom}}$ & $v_\mathrm{hom}$
 & $f_{a_s} (\%) $ & $f_{a_b} (\%) $ & $l_{s, \mathrm{hom}}$ & $l_{b, \mathrm{hom}}$ & $\frac{\tau_{s,\mathrm{hom}}}{\langle \tau \rangle}$ & $\frac{\tau_{b,\mathrm{hom}}}{\langle \tau \rangle}$ \\ \hline

Reed & 962 & 39 & 1.62
	& 23.38 & 16.03 & 18329.97 & 4.36 & 0.40 & 3.29 & 5.75 & 94.25 & 5.68 & 1.71 & 1.0 & 1.0 \\
Simmons & 1510 & 43 & 2.57
	& 25.14 & 16.64 & 36247.05 & 4.53 & 0.39 & 3.40 & 3.79 & 96.21 & 6.03 & 1.93 & 1.0 & 1.0 \\
Caltech & 762 & 43 & 1.54
	& 21.94 & 15.55 & 17083.39 & 4.25 & 0.36 & 3.13 & 7.15 & 92.85 & 5.39 & 1.54 & 1.0 & 1.0 \\
Haverford & 1446 & 82 & 1.50
	& 25.86 & 15.60 & 78135.33 & 4.31 & 0.38 & 2.98 & 2.63 & 97.37 & 6.17 & 2.46 & 1.0 & 1.0 \\
Swarthmore & 1657 & 73 & 2.32
	& 24.39 & 15.97 & 83141.94 & 4.40 & 0.39 & 3.10 & 3.12 & 96.88 & 5.88 & 1.98 & 1.0 & 1.0 \\
\hline
& & & & \multicolumn{12}{c}{Uniformly random distribution of timers} \\
 & N & z & $\langle l \rangle$
 & $T_\mathrm{unif}$ & $t_\mathrm{0.9, unif}$ & $a_\mathrm{unif}$ & $\mu_{l,\mathrm{unif}}$ & $\sigma_{l,\mathrm{unif}}$ & $v_\mathrm{unif}$
 & $f_{a_s} (\%) $ & $f_{a_b} (\%) $ & $l_{s, \mathrm{unif}}$ & $l_{b, \mathrm{unif}}$ & $\frac{\tau_{s,\mathrm{unif}}}{\langle \tau \rangle}$ & $\frac{\tau_{b,\mathrm{unif}}}{\langle \tau \rangle}$ \\ \hline

Reed & 962 & 39 & 1.62
	& 21.92 & 11.96 & 3223.54 & 5.01 & 0.60 & 4.11 & 0.95 & 99.05 & 8.00 & 2.43 & 0.28 & 1.02 \\
Simmons & 1510 & 43 & 2.57
	& 23.34 & 12.13 & 6516.15 & 5.40 & 0.57 & 4.31 & 1.30 & 98.70 & 8.19 & 2.45 & 0.24 & 1.02 \\
Caltech & 762 & 43 & 1.54
	& 21.19 & 11.85 & 2721.44 & 4.83 & 0.52 & 3.89 & 1.55 & 98.45 & 7.30 & 2.10 & 0.28 & 1.03 \\
Haverford & 1446 & 82 & 1.50
	& 22.04 & 11.50 & 13573.66 & 5.33 & 0.57 & 3.79 & 0.83 & 99.17 & 8.11 & 2.51 & 0.21 & 1.02 \\
Swarthmore & 1657 & 73 & 2.32
	& 21.02 & 11.66 & 14358.79 & 5.41 & 0.56 & 3.94 & 0.95 & 99.05 & 8.13 & 2.38 & 0.18 & 1.02 \\
\hline
& & & & \multicolumn{12}{c}{Gamma distribution (and then rounding down to an integer) of timers} \\
 & N & z & $\langle l \rangle$
 & $T_\mathrm{Gam}$ & $t_\mathrm{0.9, Gam}$ & $a_\mathrm{Gam}$ & $\mu_{l,\mathrm{Gam}}$ & $\sigma_{l,\mathrm{Gam}}$ & $v_\mathrm{Gam}$
 & $f_{a_s} (\%) $ & $f_{a_b} (\%) $ & $l_{s, \mathrm{Gam}}$ & $l_{b, \mathrm{Gam}}$ & $\frac{\tau_{s,\mathrm{Gam}}}{\langle \tau \rangle}$ & $\frac{\tau_{b,\mathrm{Gam}}}{\langle \tau \rangle}$ \\ \hline

Reed & 962 & 39 & 1.62
	& 37.60 & 15.57 & 2902.57 & 4.40 & 0.47 & 4.23 & 1.89 & 98.11 & 6.55 & 2.12 & 0.45 & 1.02 \\
Simmons & 1510 & 43 & 2.57
	& 39.84 & 15.86 & 5069.47 & 4.62 & 0.46 & 4.48 & 1.78 & 98.22 & 6.72 & 2.17 & 0.43 & 1.02 \\
Caltech & 762 & 43 & 1.54
	& 36.15 & 15.39 & 2586.5 & 4.29 & 0.44 & 4.01 & 2.32 & 97.68 & 6.20 & 1.96 & 0.45 & 1.03 \\
Haverford & 1446 & 82 & 1.50
	& 38.24 & 15.15 & 8664.29 & 4.51 & 0.43 & 4.11 & 1.03 & 98.97 & 6.88 & 2.59 & 0.47 & 1.01 \\
Swarthmore & 1657 & 73 & 2.32
	& 38.71 & 15.28 & 9306.55 & 4.58 & 0.44 & 4.24 & 1.50 & 98.50 & 6.62 & 2.19 & 0.38 & 1.02 \\
\end{tabular}
\end{ruledtabular}
\end{table*}


\section{Conclusions} \label{sec:conclusion}

In this paper, we investigated how the delay of adoption of individuals can affect the spread of adoptions in a network. We explored this idea by augmenting the Watts threshold model (WTM) with timers on the nodes, and we examined how incorporating homogeneous and heterogeneous timers can affect the dynamics of contagions. For the WTM with homogeneous timers, neither the adoption order of nodes nor adoption paths change, because a homogeneous timer simply delays the adoption of each individual by the same amount of time. However, heterogeneous timers can change both the adoption order of nodes and adoption paths, and the precise effects depend in an interesting way on both network structure and timer distribution. For example, we observed that incorporating heterogeneous timers into the WTM can either accelerate or decelerate the time to a steady state in comparison to incorporating a homogeneous timer. However, in our calculations using two different types of heterogeneous timer distributions (specifically, uniformly random timers and timers determined using a Gamma distribution and then rounding down to an integer), we found that the majority of nodes (up to $90\%$ in our study) in networks adopt earlier with homogeneous timers than with heterogeneous ones. We speculate that stems --- the longest adoption paths in a dissemination tree --- play a significant role in spreading adoptions faster to the majority of nodes in networks for the WTM with heterogeneous timers than for the WTM with a homogeneous timer. To support our speculation, we examined the delay of adoption processes when changing the timers of stem nodes to the mean value of timers, and we found that adoption processes are delayed even though these nodes constitute fewer than $1\%$ of the adopted nodes in the examined networks. We also developed a pair approximation for the WTM with timers that gives good agreement with our numerical computations for the temporal evolution of the fraction of adopted nodes. 

In future work, we seek to investigate real-world data for the spread of information, memes, innovations, misinformation, and other phenomena on networks using models that include latencies (i.e., timers) in addition to influence thresholds. People can be late adopters for a variety of reasons (e.g., hard to convince versus easy to convince but lazy), and incorporating timers into contagion models should be helpful for exploring different mechanisms of late adoption and their possible consequences for forecasting virality. In these studies, it will be important to augment a variety of different spreading models with timers. The deterministic update rule in the WTM made it particularly suitable for a first foray into incorporating timers in spreading processes, and that is why we used it in the present paper. It will be interesting to incorporate timers into stochastic spreading processes to examine how this changes spreading behavior and virality forecasts.



\begin{acknowledgements}

We thank Javier Borge-Holthoefer, Jacob Fisher, James Gleeson, Peter Grindrod, Anne Kandler, Beom-Jun Kim, Mikko Kivel\"a, Florian Klimm, Sang Hoon Lee, James Moody, Chinami Oka, Tim Rogers, Seung-Woo Son, and Felix Reed-Tsochas for very helpful conversations. S-WO is supported by a Kwanjeong Fellowship from the Kwanjeong Educational Foundation.

\end{acknowledgements}


\appendix

\section{Adoption paths when nodes have large adoption thresholds} \label{app:large_threshold}

\begin{figure}[H]
  \centering
		\includegraphics[width=0.275\textwidth]{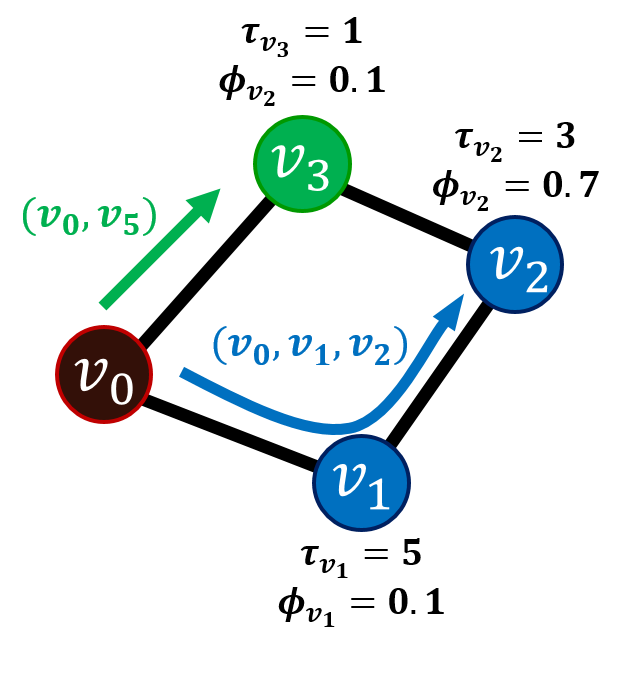}
	\caption{Example of adoption paths when there is a node with a large adoption threshold.}  
\label{fig:large_threshold}
\end{figure}

When most node timers are small, we saw in Section~\ref{sec} that we tend to obtain long adoption paths only when the node adoption thresholds are sufficiently small (e.g., so that only one neighbor needs to adopt for a node to adopt). In this appendix, we consider adoption paths when some nodes have large adoption thresholds. In Fig.~\ref{fig:large_threshold}, we illustrate an example network in which the last node to adopt ($v_2$ in this case) has a sufficiently large threshold to adopt after both nodes $v_1$ and $v_3$ adopt. Therefore, when the timers of $v_1$ and $v_3$ are different, the node that adopts later triggers the adoption of node $v_2$ and becomes part of the longer adoption path. In our example in Fig.~\ref{fig:large_threshold}, $v_2$ has a larger timer than $v_3$, so $v_2$ yields a longer adoption path [given by $(v_0, v_1, v_2)$] than $v_3$. Accordingly, if we abandon the condition that a node adopts if one neighbor adopts, nodes with large timers can become part of long adoption paths.

\section{Stems with large timers} \label{app:exceptional_case}

\begin{figure}[H]
  \centering
		\includegraphics[width=0.5\textwidth]{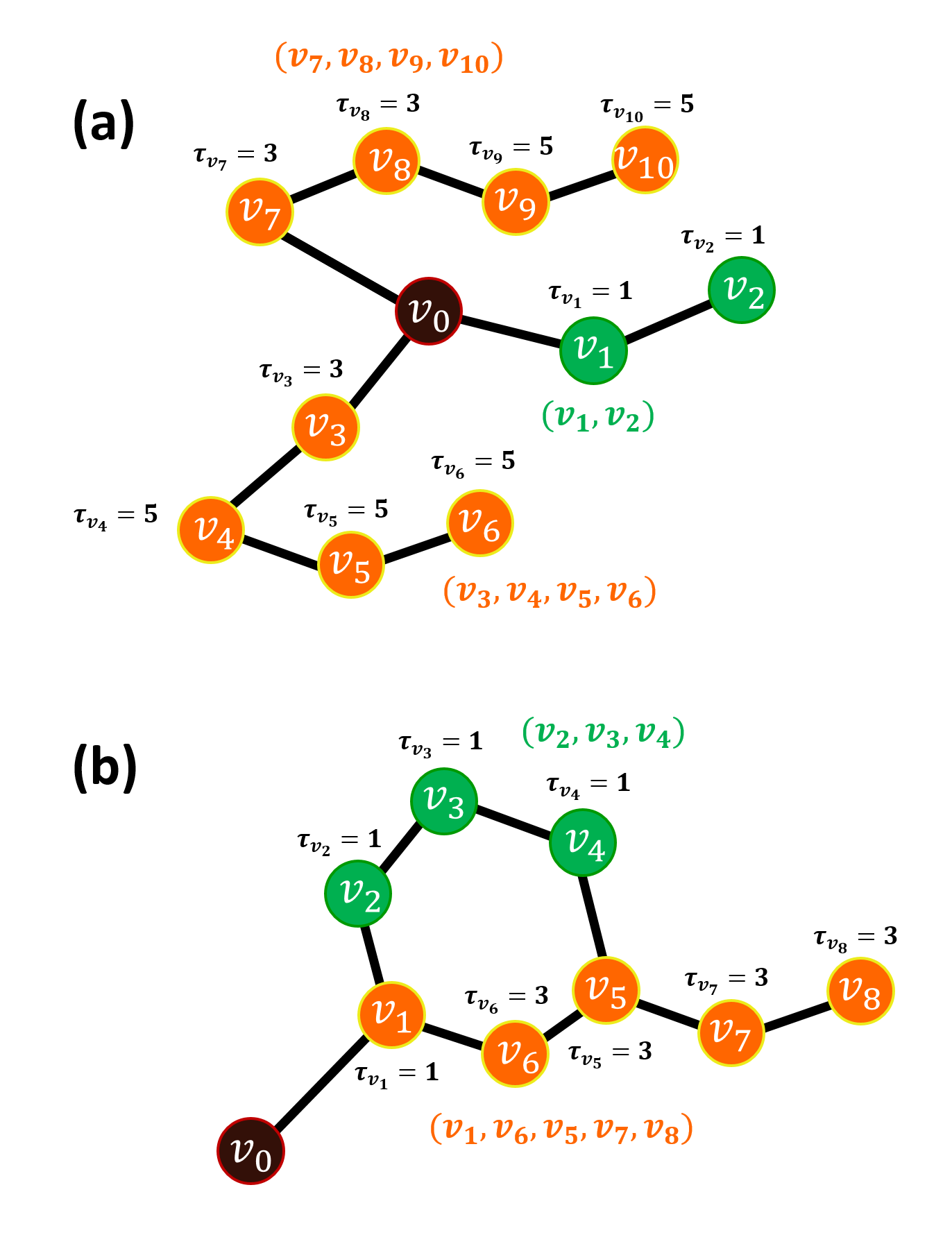}
	\caption{Examples of networks with larger mean stem-node timers than mean branch-node timers.  
	}
\label{fig:exceptional_case}
\end{figure}

In this appendix, we discuss situations in which a stem in a network includes nodes with large timers, in contrast to those that we investigated in Section~\ref{sec:adoption_path_in_large_networks}. The characteristics of stems and branches can depend on network structure and the timers. Previously, we focused on examining homogeneous adoption thresholds such that nodes with degree equal to a network's mean degree meet the threshold adoption condition if even a single neighbor has adopted. We thus observed that stem nodes tend to have mean timers smaller than the mean branch-node timers. However, depending on network structure (see Fig.~\ref{fig:exceptional_case}a) and the assignments of timers (see Fig.~\ref{fig:exceptional_case}b), it is possible for the mean stem-node timers to be larger than the mean branch-node timers. In Fig.~\ref{fig:exceptional_case}a, the adoption paths do not change, regardless of the assignments of thresholds and timers, because the seed node is not adjacent to any node in a cycle of length $4$ or more (See Section~\ref{sec:timer_model_on_small_networks}). Therefore, in Fig.~\ref{fig:exceptional_case}a, the stems $(v_3, v_4, v_5, v_6)$ and $(v_7,v_8,v_9,v_{10})$ have larger mean timers than the branches. (Note that the number of stems is larger than the number of branches, which is also different from our findings on large random networks in Table~\ref{tab:adoption_path_random_networks} and \ref{tab:adoption_path_real_networks}.)

Even if a network includes a cycle with $4$ or more nodes, the stem nodes can have a larger mean timer than the branch nodes, depending on network structure and timer assignments. With the given timer assignment, the network in Fig.~\ref{fig:exceptional_case}b has the stem $(v_1,v_6,v_5,v_7,v_8)$, which has a larger mean timer than the branch $(v_2,v_3,v_4)$.

\section{Cycles in dissemination trees} \label{app:cycle}

For a 3-cycle to exist in a dissemination tree's underlying undirected graph, a node's timer needs to be triggered by the simultaneous adoption of its two neighbors. However, the two neighbors cannot be adjacent to each other in the dissemination tree because any two adjacent nodes in a dissemination tree cannot adopt simultaneously; a pair of nodes become neighbors in a dissemination tree if and only if one triggers the timer of the other. Therefore, there cannot exist a 3-cycle in a dissemination tree.

\begin{figure}[H]
  \centering
	\includegraphics[width=0.5\textwidth]{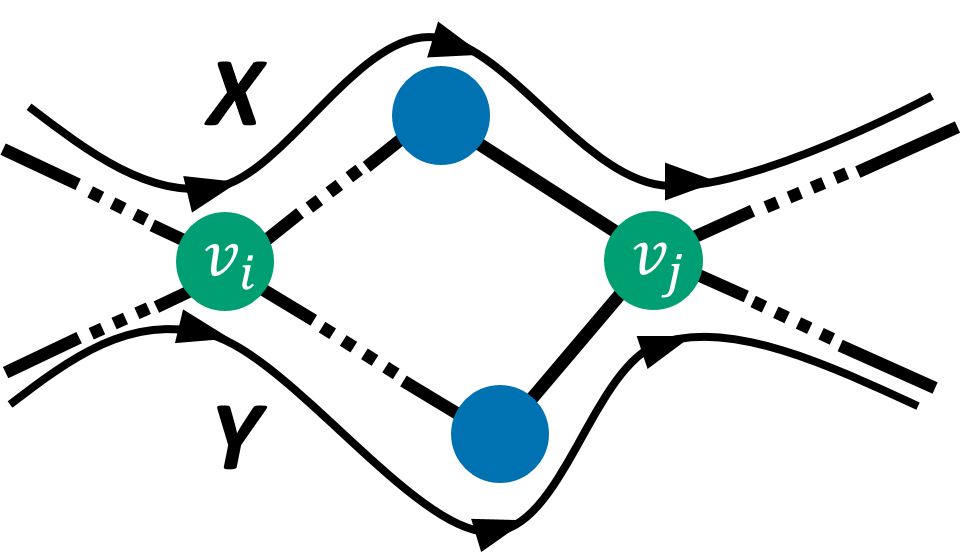}
	\caption{Illustration of a cycle in a dissemination tree.}
\label{fig:cycle}
\end{figure}

In contrast, a dissemination tree can have cycles with 4 or more nodes when two nodes have a common predecessor in each of their adoption paths. In particular, for the WTM with heterogeneous timers, any $k$-cycle with $k \geq 4$ can exist; for a homogeneous timer, $k$ must be even. Assume that we have a cycle in a dissemination tree (see Fig.~\ref{fig:cycle}). For a dissemination tree, having a cycle implies that there exist two different adoption paths ($X$ and $Y$ in Fig.~\ref{fig:cycle}) with at least two common nodes ($v_i$ and $v_j$) in their adoption paths. Suppose that $v_i$ adopts before $v_j$. If the timers are homogeneous, the adoption spreads from $v_i$ to $v_j$ at the same rate in $X$ and $Y$. Therefore, the cycle sizes for a homogeneous timer can only be even.
 However, for heterogeneous timers, the rates of adoption spreading in $X$ and $Y$ can be different. Therefore, any $k$-cycle with $k \geq 4$ can exist for the WTM with heterogeneous timers.

\section{Time to reach certain fractions of adopted nodes for {\sc Facebook100} networks}\label{app:slower_stem_on_other_random_networks}

As in Section~\ref{sec:dissemination_tree}, we change the timers of stem nodes of dissemination trees to the mean value $\langle \tau \rangle$ of the timers (which is $\mu_{\tau}=4$ in our example). The times that the stem nodes' neighbors in a tree adopt then also change, as the adoption time of a node in an adoption path is determined by the sum of its timer and the adoption time of its predecessor node. (Note that we are not rerunning the WTM dynamics.)

\begin{figure*}[t!]
	\begin{subfigure}[b]{0.495\textwidth}
		\includegraphics[width=\textwidth]{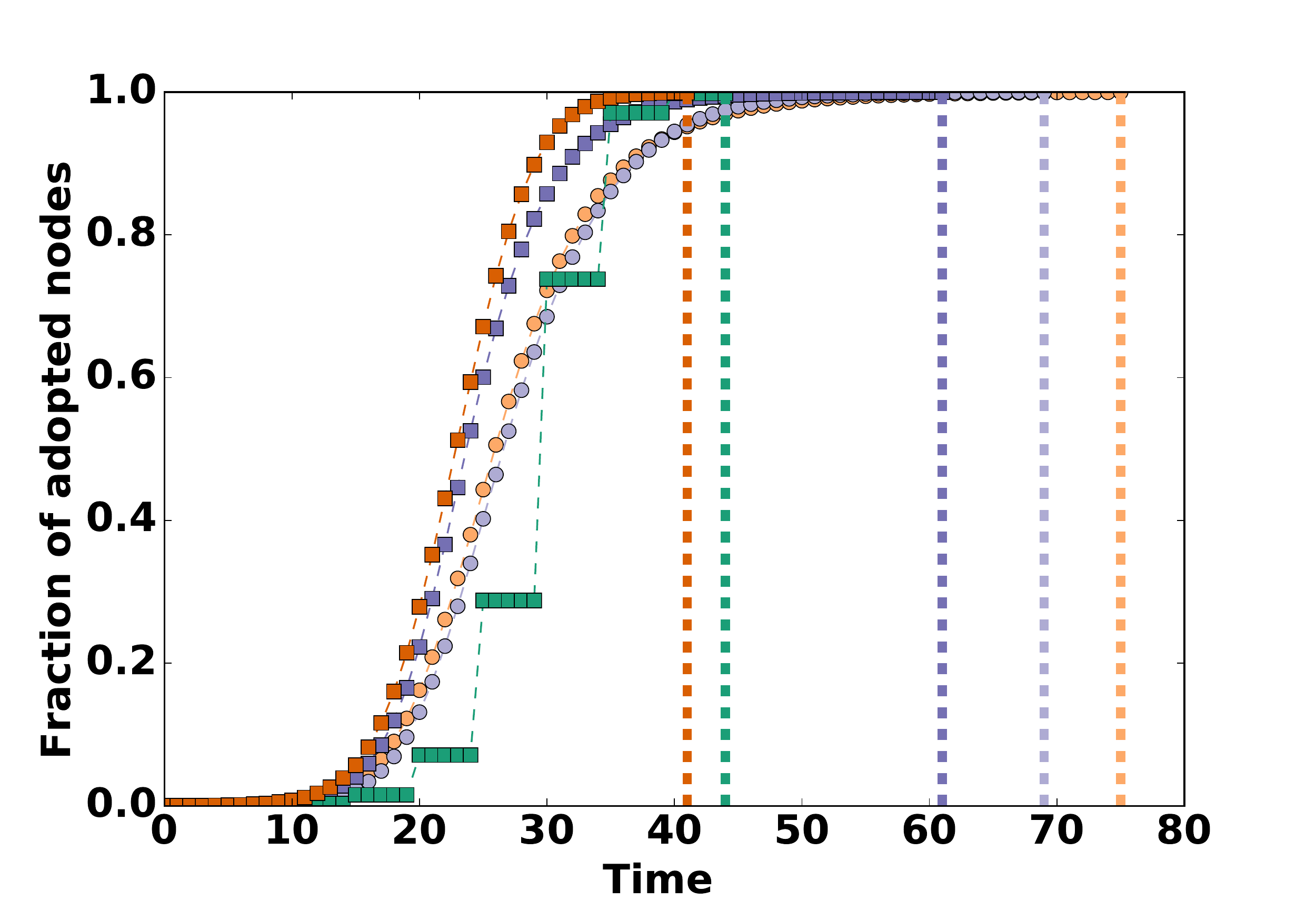}
		\caption{$\alpha=0.5$}
		\label{fig:clique_3_5}
	\end{subfigure}
	\begin{subfigure}[b]{0.495\textwidth}
		\includegraphics[width=\textwidth]{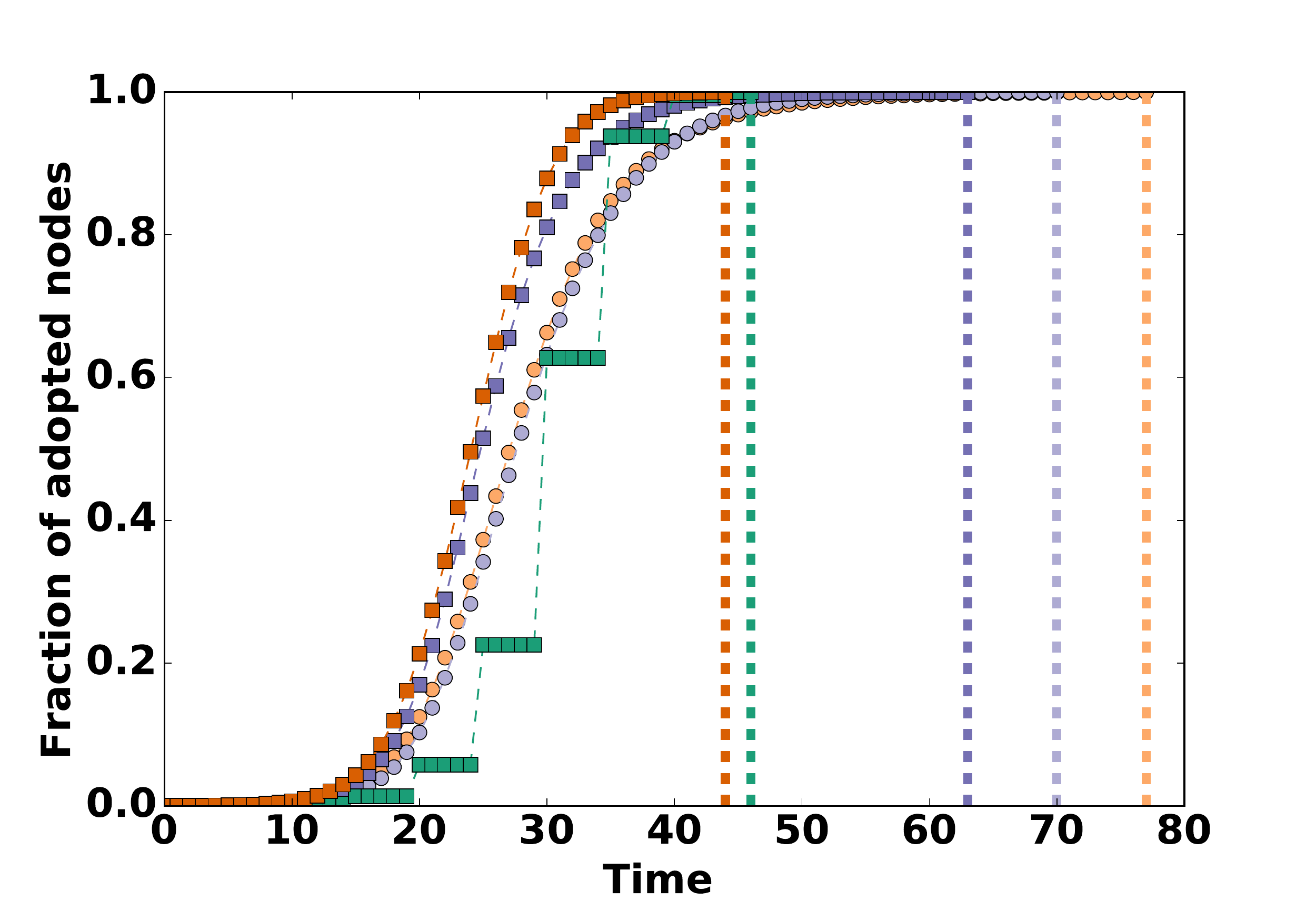}
		\caption{$\alpha=1$}
		\label{fig:clique_3_1}
	\end{subfigure}
	\begin{subfigure}[b]{0.495\textwidth}
		\includegraphics[width=\textwidth]{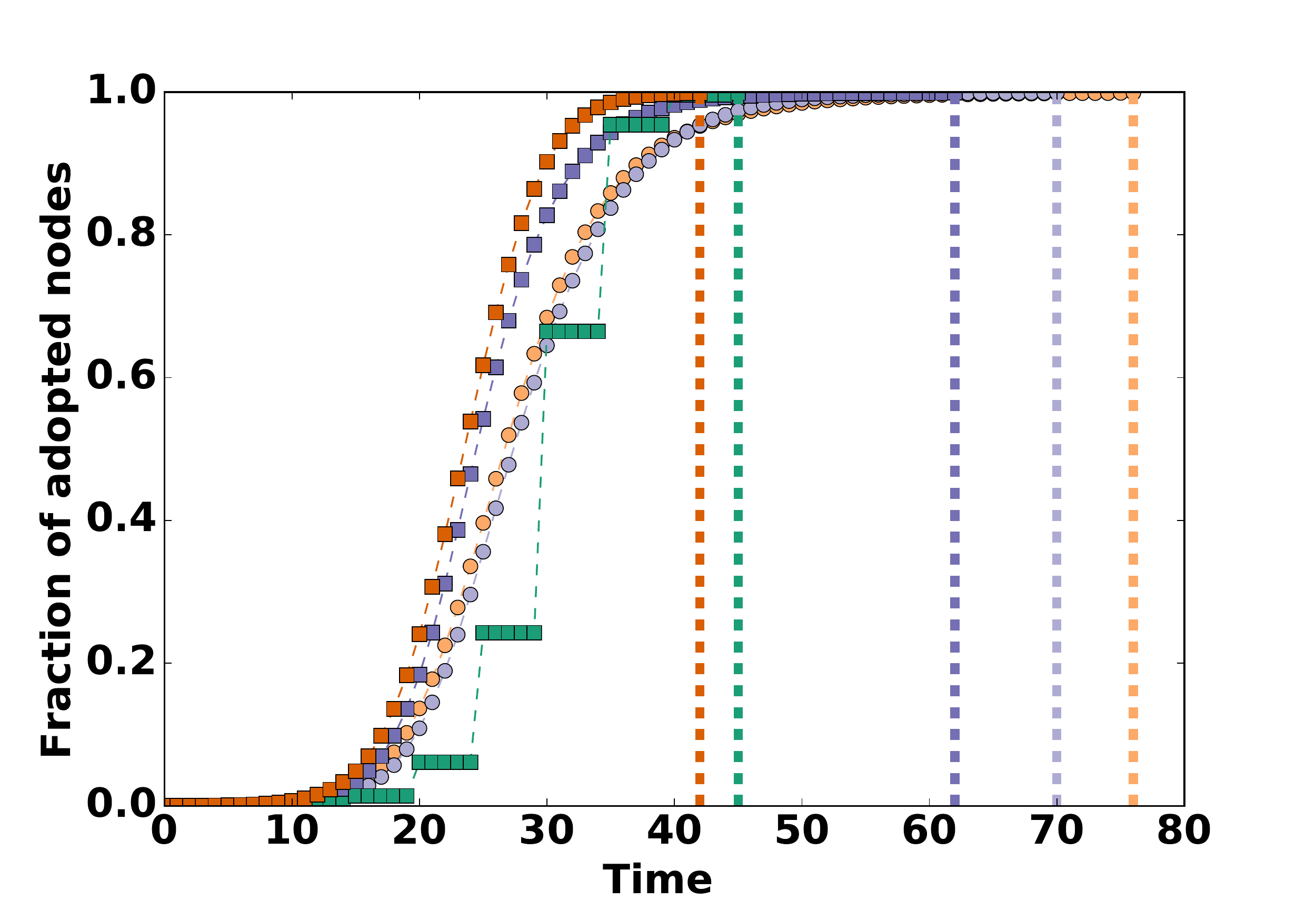}
		\caption{$\beta=0.5$}
		\label{fig:clique_4_5}
	\end{subfigure}
	\begin{subfigure}[b]{0.495\textwidth}
		\includegraphics[width=\textwidth]{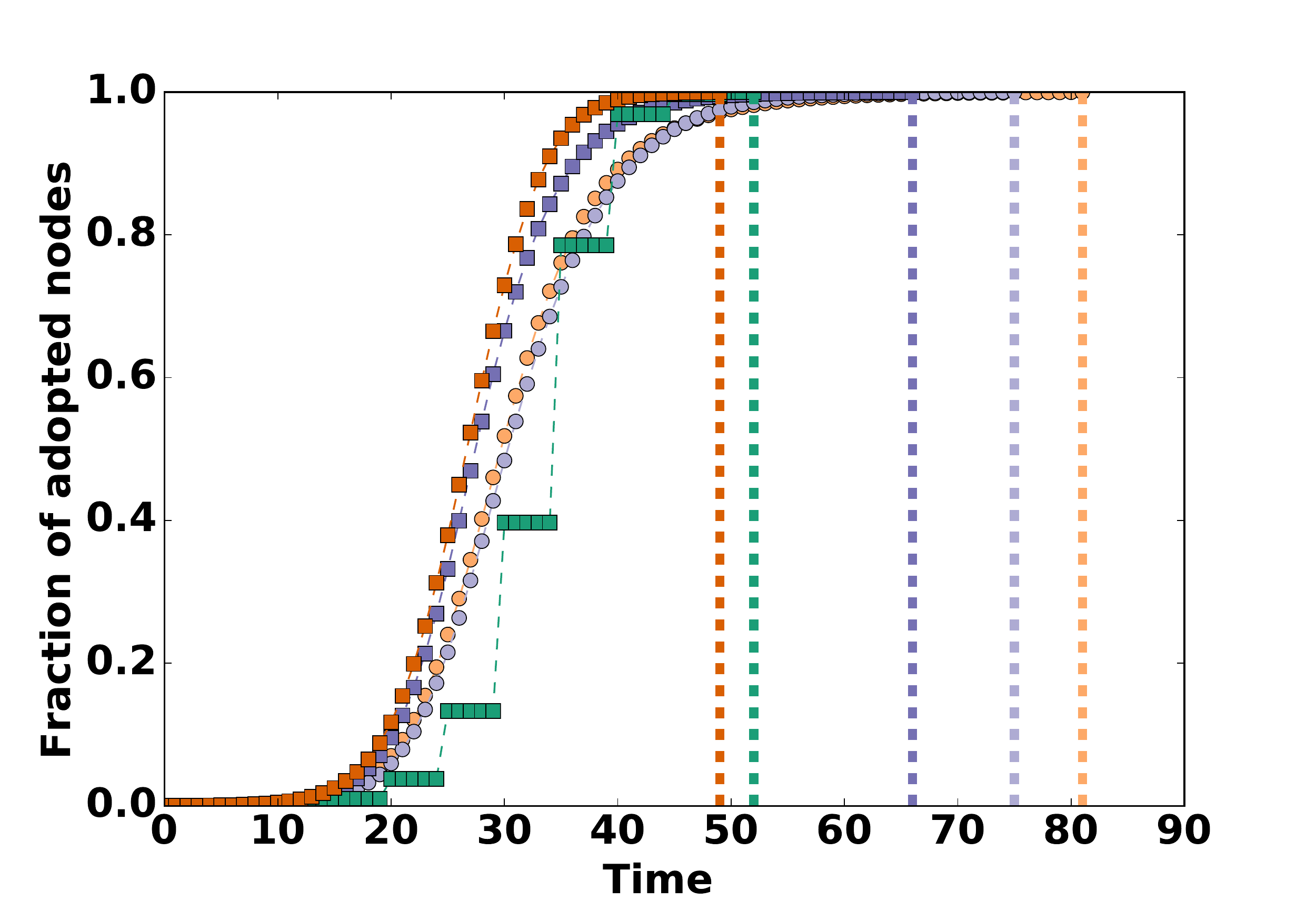}
		\caption{$\beta=1$}
		\label{fig:clique_4_1}
	\end{subfigure}
	\caption{\raggedright Adoption curves of the WTM with timers on generalized configuration-model networks with $N=10,000$ nodes and a Poisson degree distribution with mean $z=6$ and adoption curves after we adjust the timers of stem nodes of dissemination trees. We use (a, b) generalized configuration-model networks with $3$-cliques and (c, d) generalized configuration-model networks with $4$-cliques. We consider different values of the edge--clique ratios $\alpha$ and $\beta$, where $\alpha$ determines the edge--clique ratio for a node of degree $k \geq 3$ in Congen-3 networks and $\beta$ determines the edge--clique ratio for a node of degree $k \geq 4$ in Congen-4 networks. We use the parameter values $\alpha=0.5$ in panel (a), $\alpha=1$ in panel (b), $\beta=0.5$ in panel (c), and $\beta=1$ in panel (d). For all networks, each node has a homogeneous threshold value of $\phi=0.1$. 
 The green curve is for the WTM with homogeneous timers with $\tau=4$, the orange curves are for the WTM with heterogeneous timers $\tau$ that are distributed uniformly at random from the set $\{0, 1, \dots, 8\}$, and the blue curves are for heterogeneous timers determined from the Gamma distribution with mean $\mu_\mathrm{\tau}=4$ and standard deviation $\sigma_{\tau}=4$ and then rounding down to an integer. To isolate the effects of different distributions of timers, in each case, we run the WTM using the same networks with the same seed nodes. The dark orange and blue curves are before we change the timers in dissemination trees, and the corresponding light-colored curves are after we change those timer values (see the main manuscript). The squares (for the dark curves) and disks (for the light curves) are the results of numerical simulations, and the dashed lines mark the times at which the adoption process of the corresponding color reaches a steady state. Our numerical results are means over $1,000$ realizations.
For each realization, we generate an independent network. We also determine the seed node (uniformly at random) and timer values separately for each realization.
 }
\label{fig:dissem_tree_change}
\end{figure*}

In Fig.~\ref{fig:dissem_tree_change}, we investigate how the adoption curves change because of the change of timers of stem nodes. We use generalized configuration-model networks with $3$-cliques (Congen-$3$) and a generalized configuration-model networks with $4$-cliques (Congen-$4$), and we consider different values of the edge--clique ratios $\alpha$ and $\beta$, where $\alpha$ determines the edge--clique ratio for a node of degree $k \geq 3$ in Congen-$3$ and $\beta$ determines the edge--clique ratio for a node of degree $k \geq 4$ in Congen-$4$. We use $\alpha=0.5$ in Fig.~\ref{fig:clique_3_5}, $\alpha=1.0$ in Fig.~\ref{fig:clique_3_1}, $\beta=0.5$ in Fig.~\ref{fig:clique_4_5}, and $\beta=1.0$ in Fig.~\ref{fig:clique_4_1}. In each figure, the orange and blue curves both represent adoption curves for the WTM with heterogeneous timers. For the orange curve, timers are distributed uniformly at random from the set $\{0, 1, \dots, 8\}$; for the blue curve, timers are determined from a Gamma distribution with mean $\mu_\mathrm{\tau}=4$ and standard deviation $\sigma_{\tau}=4$ and are then rounded down to an integer. The curves with lighter colors are the adoption curves after we change the timers of stem nodes of dissemination trees. The light orange curve is for uniformly-randomly-distributed timers, and the light blue curve is for Gamma-distributed timers. The green curve, which we include for comparison, is for a WTM with a homogeneous timer. We observe that the light-colored curves intersect the green curve earlier than do the dark-colored curves.

In Table~\ref{tab:dissem_tree_change}, we show the change in times to achieve certain fractions $\rho^*$ of nodes --- calculated with Eq.~(6) from the main text --- after adjusting the timers of stem nodes. From both Fig.~\ref{fig:dissem_tree_change} and Table~\ref{tab:dissem_tree_change}, we observe that adoption processes for heterogeneous timers are delayed by changing the timers of stem nodes, which constitute fewer than $1\%$ of all nodes in the networks.

\begin{table*}
\caption{\label{tab:dissem_tree_change}
\raggedright Time to reach certain fractions of adopted nodes for the WTM with timers on generalized configuration-model networks with $3$-cliques (Congen-$3$) and generalized configuration-model networks with $4$-cliques (Congen-$4$) and the change of time to reach certain fractions of adopted nodes after we adjust the timers of stem nodes of dissemination tree to the mean value $\mu_\tau$ of all timers. We consider different values of the edge--clique ratios $\alpha$ and $\beta$, where $\alpha$ determines the edge--clique ratio for a node of degree $k \geq 3$ in Congen-$3$ and $\beta$ determines the edge--clique ratio for a node of degree $k \geq 4$ in Congen-$4$. For all networks in this table, we start with configuration-model networks with $N=10,000$ nodes and degrees drawn from a Poisson distribution $P_k = {z^k e^{-z}}/{k!}$ with mean $z=6$. All nodes have a homogeneous threshold of $\phi=0.1$, so a node with the mean degree adopts once one of its neighbors adopts. We consider a homogeneous timer with $\tau=4$, heterogeneous timers distributed uniformly at random from the set $\{0, 1, \dots, 8\}$, and timers determined using a Gamma distribution with mean $\mu_\tau=4$ and standard deviation $\sigma_{\tau}=4$ and then rounded down to an integer. To isolate the effects of different distributions of timers, in each case, we run the WTM using the same networks with the same seed nodes. Our numerical results are means over $1,000$ realizations. For each realization, we generate an independent network. We also determine the seed node (uniformly at random) and timer values separately for each realization.}
\centering
\begin{ruledtabular}
\begin{tabular}{c|||c|cc|cc||c|cc|cc}
& \multicolumn{10}{c}{Congen-$3$} \\
& \multicolumn{5}{c||}{$\alpha=0.5$} & \multicolumn{5}{c}{$\alpha=1.0$} \\
\hline
$\rho^*$\footnotemark[1] & $t_\mathrm{hom}$\footnotemark[2]
& $t_\mathrm{unif,WTM}$\footnotemark[3] & $t_\mathrm{unif,dis}$\footnotemark[4] & $t_\mathrm{Gam,WTM}$\footnotemark[5] & $t_\mathrm{Gam,dis}$\footnotemark[6]
& $t_\mathrm{hom}$\footnotemark[2] & $t_\mathrm{unif,WTM}$\footnotemark[3] & $t_\mathrm{unif,dis}$\footnotemark[4] & $t_\mathrm{Gam,WTM}$\footnotemark[5] & $t_\mathrm{Gam,dis}$\footnotemark[6] \\
\hline
0.5 & 29.82 & 22.94 & 26.05 & 23.71 & 26.63 & 30.56 & 24.17 & 27.18 & 24.87 & 27.67 \\
0.6 & 30.43 & 24.05 & 27.37 & 24.87 & 28.19 & 31.52 & 25.32 & 28.76 & 26.07 & 29.26 \\
0.7 & 31.27 & 25.23 & 29.37 & 26.22 & 30.02 & 32.70 & 26.53 & 30.58 & 27.46 & 31.19 \\
0.8 & 32.53 & 26.56 & 31.67 & 27.99 & 32.37 & 34.25 & 27.91 & 32.95 & 29.27 & 33.49 \\
0.9 & 34.30 & 28.34 & 35.47 & 30.93 & 36.07 & 35.43 & 29.77 & 36.88 & 32.23 & 37.49 \\
1.0 & 44.06 & 41.76 & 75.70 & 61.92 & 69.01 & 46.32 & 44.10 & 77.67 & 63.67 & 70.87 \\

\hline
\hline
& \multicolumn{10}{c}{Congen-$4$} \\
& \multicolumn{5}{c||}{$\beta=0.5$} & \multicolumn{5}{c}{$\beta=1.0$} \\
\hline
$\rho^*$\footnotemark[1] & $t_\mathrm{hom}$\footnotemark[2]
& $t_\mathrm{unif,WTM}$\footnotemark[3] & $t_\mathrm{unif,dis}$\footnotemark[4] & $t_\mathrm{Gam,WTM}$\footnotemark[5] & $t_\mathrm{Gam,dis}$\footnotemark[6]
& $t_\mathrm{hom}$\footnotemark[2] & $t_\mathrm{unif,WTM}$\footnotemark[3] & $t_\mathrm{unif,dis}$\footnotemark[4] & $t_\mathrm{Gam,WTM}$\footnotemark[5] & $t_\mathrm{Gam,dis}$\footnotemark[6] \\
\hline
0.5 & 30.26 & 23.60 & 26.75 & 24.47 & 27.39 & 33.57 & 26.81 & 29.81 & 27.58 & 30.38 \\
0.6 & 31.08 & 24.74 & 28.32 & 25.65 & 28.96 & 34.71 & 28.04 & 31.45 & 28.87 & 32.05 \\
0.7 & 32.14 & 25.93 & 30.12 & 27.02 & 30.80 & 35.69 & 29.33 & 33.32 & 30.33 & 33.98 \\
0.8 & 33.62 & 27.29 & 32.46 & 28.81 & 33.18 & 36.83 & 30.81 & 35.72 & 32.21 & 36.46 \\
0.9 & 35.05 & 29.11 & 36.35 & 31.75 & 36.91 & 38.78 & 32.81 & 39.68 & 35.23 & 40.35 \\
1.0 & 45.43 & 42.79 & 76.63 & 62.86 & 70.14 & 52.17 & 49.60 & 81.77 & 66.99 & 75.53 \\
\end{tabular}
\end{ruledtabular}
\footnotetext[1]{Fraction of adopted nodes}
\footnotetext[2]{Time to reach $\rho^*$ for the WTM with a homogeneous timer.}
\footnotetext[3]{Time to reach $\rho^*$ for the WTM with timers distributed uniformly at random.}
\footnotetext[4]{Time to reach $\rho^*$ after we adjust the timers of stem nodes for timers distributed uniformly at random.}
\footnotetext[5]{Time to reach $\rho^*$ for the WTM with timers determined using a Gamma distribution and then rounding down to an integer.}
\footnotetext[6]{Time to reach $\rho^*$ after we adjust the timers of stem nodes for timers determined using a Gamma distribution and then rounding down to an integer.}
\end{table*}


\section{Time to reach certain fractions of adopted nodes for {\sc Facebook100} networks}
\label{app:real_net_adoption_percentage}

In Table~\ref{tab:time_to_certain_percent_real_networks}, we give the times to reach certain fractions $\rho^*$ of adopted nodes for {\sc Facebook100} networks when timers are homogeneous, distributed uniformly at random, and determined using a Gamma distribution and then rounded down to an integer.

\begin{table*}
\caption{\label{tab:time_to_certain_percent_real_networks}
\raggedright Time to reach certain fractions of adopted nodes in the WTM model with homogeneous and heterogeneous timers on {\sc Facebook100} networks. The quantity $N$ is the number of nodes in a network, $z$ is its mean degree, and $\langle l \rangle$ is its mean shortest-path length between pairs of nodes. All nodes have a homogeneous adoption threshold of $\phi=0.01$.
The notation $x_\mathrm{hom}$ indicates that we calculate the quantity $x$ for homogeneous timers, the notation $x_\mathrm{unif}$ indicates that we calculate the quantity $x$ for timers that are are distributed uniformly at random, and the notation $x_\mathrm{Gam}$ indicates that we calculate the quantity $x$ for timers determined using a Gamma distribution and then rounded down to an integer. The quantity $T$ is the time that it takes to reach a steady state, $t_\mathrm{\rho^*}$ is the time that it takes for the fraction $\rho^*$ of adopted nodes to reach $\rho^*$. 
Each quantity is a mean over $100$ realizations with different distribution of timers for each realization, and we also choose the seed node uniformly at random for each realization. 
}
\begin{ruledtabular}
\begin{tabular}{p{2cm}||ccc|cccccc}
& & & & \multicolumn{6}{c}{Homogeneous timer} \\

 & N & z & $\langle l \rangle$
 & $T_\mathrm{hom}$ & $t_\mathrm{0.9, hom}$ & $t_\mathrm{0.8, hom}$ & $t_\mathrm{0.7, hom}$ & $t_\mathrm{0.6, hom}$ & $t_\mathrm{0.5, hom}$ \\ \hline

Reed & 962 & 39 & 1.62
	& 23.40 & 16.19 & 15.22 & 14.40 & 13.44 & 12.55 \\
Simmons & 1510 & 43 & 2.57
	& 25.34 & 16.90 & 15.71 & 15.27 & 14.44 & 13.69 \\
Caltech & 762 & 43 & 1.54
	& 21.95 & 15.60 & 14.75 & 14.01 & 12.91 & 12.07 \\
Haverford & 1446 & 82 & 1.50
	& 25.83 & 15.38 & 14.96 & 14.51 & 13.71 & 13.17 \\
Swarthmore & 1657 & 73 & 2.32
	& 24.35 & 15.91 & 15.17 & 14.69 & 14.20 & 13.37 \\
\hline
\hline
& & & & \multicolumn{6}{c}{Timers distributed uniformly at random} \\
 & N & z & $\langle l \rangle$
 & $T_\mathrm{unif}$ & $t_\mathrm{0.9, unif}$ & $t_\mathrm{0.8, unif}$ & $t_\mathrm{0.7, unif}$ & $t_\mathrm{0.6, unif}$ & $t_\mathrm{0.5, unif}$ \\ \hline

Reed & 962 & 39 & 1.62
	& 22.0 & 12.08 & 10.92 & 9.96 & 9.03 & 8.10 \\
Simmons & 1510 & 43 & 2.57
	& 23.47 & 12.22 & 11.07 & 10.11 & 9.18 & 8.26 \\
Caltech & 762 & 43 & 1.54
	& 21.24 & 11.84 & 10.69 & 9.72 & 8.80 & 7.86 \\
Haverford & 1446 & 82 & 1.50
	& 22.36 & 11.45 & 10.43 & 9.50 & 8.59 & 7.68 \\
Swarthmore & 1657 & 73 & 2.32
	& 21.01 & 11.61 & 10.58 & 9.65 & 8.73 & 7.82 \\
\hline
\hline
& & & & \multicolumn{6}{c}{Timers determined from a Gamma distribution and then rounding down to an integer} \\
 & N & z & $\langle l \rangle$
 & $T_\mathrm{Gam}$ & $t_\mathrm{0.9, Gam}$ & $t_\mathrm{0.8, Gam}$ & $t_\mathrm{0.7, Gam}$ & $t_\mathrm{0.6, Gam}$ & $t_\mathrm{0.5, Gam}$ \\ \hline

Reed & 962 & 39 & 1.62
	& 37.33 & 15.58 & 12.63 & 10.90 & 9.66 & 8.66 \\
Simmons & 1510 & 43 & 2.57
	& 39.92 & 15.95 & 13.02 & 11.30 & 10.06 & 9.07 \\
Caltech & 762 & 43 & 1.54
	& 37.04 & 15.58 & 13.20 & 11.42 & 10.15 & 9.15 \\
Haverford & 1446 & 82 & 1.50
	& 38.44 & 15.12 & 12.28 & 10.62 & 9.44 & 8.50 \\
Swarthmore & 1657 & 73 & 2.32
	& 38.63 & 15.27 & 12.45 & 10.79 & 9.60 & 8.68 \\
\end{tabular}
\end{ruledtabular}
\end{table*}




\end{document}